\documentclass[
aps,
prd,
showkeys,
showpacs,
superscriptaddress,
groupedaddress,
amsfonts
]{revtex4-2}
\usepackage{amsmath}
\usepackage{multido}
\usepackage{romannum}
\makeatletter
\ams@newcommand{\vardot}[2]{%
	{\mathop{#2\kern0pt}\limits^{\vbox to-1.4\ex@{\kern-\tw@\ex@
				\hbox{\normalfont\multido{}{#1}{.}}\vss}}}}
\makeatother
\newtheorem{definition}{Definition}[section]
\usepackage{hyperref}
\usepackage{amsmath,amsfonts,amssymb}
\hypersetup{dvips,dvipdfm,colorlinks=true,urlcolor=green,filecolor=magenta,linktoc=page,citecolor=red,linkcolor=blue,bookmarks=true}
\usepackage{graphicx,epstopdf}
\usepackage{dcolumn}   
\usepackage{subcaption}
\usepackage{multirow}
\usepackage{multirow,booktabs}
\usepackage{makecell}
\usepackage[english]{babel}
\newtheorem{theorem}{Theorem}
\usepackage{appendix}
\newtheorem{prop}{Proposition}

\begin{document}

\title{ Dynamical system analysis of quintessence dark energy model}

\author{Soumya Chakraborty}
\email{soumyachakraborty150@gmail.com; soumyac.math.rs@jadavpuruniversity.in}
\affiliation{Department of Mathematics, Jadavpur University, Kolkata 700032, West Bengal, India}
\author{Sudip Mishra}
\email[Corresponding author:]{sudipcmiiitmath@gmail.com; sudip.mishra@makautwb.ac.in}
\affiliation{Department of Applied Mathematics, Maulana Abul Kalam Azad University of Technology,
	Kolkata-700064 and Haringhata, Nadia-741249 (main campus), West Bengal, India}
\author{Subenoy Chakraborty}
\email{schakraborty.math@gmail.com}
\affiliation{Department of Mathematics, Jadavpur University, Kolkata-700 032, West Bengal, India}

\begin{abstract}
	
Our work deals with the dynamical system analysis of quintessence dark energy scalar field model with exponential potential.  A dynamical system analysis has been applied at the background level.  Using suitable transformation of variables, the evolution equations are reduced to an autonomous system for exponential form of the scalar potential.  The critical points are analyzed with center manifold theory and stability has been discussed by using Schwarzian derivative.  Finally, cosmological implications of the critical points are discussed and it is found that the stability of the late-time attractor changes for quintessence dark energy model.
	
\end{abstract}

\keywords{Quintessence dark energy model, Discrete dynamical system, Fixed points, Stability, Schwarzian derivative, Center manifold theory.}
	
	\pacs{98.80.Cq, 98.80.-k, 95.35.+d, 95.36.+x, 05.70.Fh, 05.45.-a, 02.10.Ud, 02.40.Vh}
	
	\maketitle	
\newpage
\section{Introduction}	
In cosmology, till the early 90's the challenging issue was to find the analytic solution of the evolution equations which are highly non-linear and coupled in nature.  As a result, it was very hard to find any cosmological inferences from the cosmological models.  But the situation changes since late 90's \cite{BAHAMONDE20181} when the dynamical system approach has been applied in the field of cosmology.  Dynamical system analysis is a very powerful mathematical tool that provides information from the evolution equations without any reference to initial conditions or any specific behavior at any intermediate instant \cite{wainwright_ellis_1997}.  There may be infinite possible evolution for a general cosmological scenario but its asymptotic behavior particularly at late times are limited to a few different classes.  These few classes can be identified as stable critical points if the cosmic evolution equations can be converted into an autonomous form. Thus by analyzing such critical points, one may infer about the late time evolution of the universe without going for any analytic solution or ambiguity to the initial conditions not arise.  So far most of the dynamical analysis at cosmological scenarios is restricted to the background level, i.e, formation of autonomous system, determination of critical points, and estimation of the relevant cosmological parameters namely density parameter, the equation of state parameter and so on.\par

The present work deals with a standard cosmological model in the context of the present accelerating phase namely the quintessence dark energy scalar field model having exponential potential.  Using suitable choice of the variables the evolution equations are converted into a discrete type autonomous system and the critical points are analyzed using center manifold theory and stability analysis has been presented with Schwarzian derivative. The manuscript is organized as follows: In Section \ref{BQS} we discuss the background of Quintessence scenarios under flat FLRW space-time.  In Section \ref{BES} we construct the autonomous system corresponding to the basic equations of the cosmological model and critical points are determined in this section. Stability analysis of all critical points for various choices of the involving parameters is shown in Section \ref{III} from the perspective of discrete dynamical system analysis.  We present global dynamical analysis and 
cosmological implications in Section \ref{IV}.  Finally, a brief discussion and important concluding remarks of this work are proposed in Sec. \ref{V}.
		
	\section{Background of Quintessence Scenarios} \label{BQS}
	
	In the perspective of phase space analysis of the basic dynamical dark energy scenario, namely the quintessence one with an exponential potential, which is the archetype quintessence scenario due to the well-posed theoretical justification of exponential potentials \cite{Sangwan:2018zpz,Liu:2022yao,Bouali:2023ncq,article,Goswami:2019zto}.  In the background of flat FLRW space-time, the Friedman equations for the general cosmological model are (choosing $8\pi G=1=c$)
 \begin{align}
     3H^2=\rho_m+\rho_d\label{equ1}
 \end{align}
	and
\begin{align}\label{FME2}
    2\dot{H}=-\left\{(\rho_m+p_m)+(\rho_d+p_d)\right\}
\end{align}

with $(\rho_m,p_m)$ and $(\rho_d,p_d)$ the energy density and thermodynamic pressure for the matter component and the dark energy component respectively. 
 The `dot' denotes the derivative with respect to time, $H=\frac{\dot{a}}{a}$ is the Hubble parameter.  The energy-momentum conservation relations for these matter components are (assuming non-interacting)

\begin{align}
    \dot{\rho_m}+3H(\rho_m+p_m)=0, \\ 
    \dot{\rho_d}+3H(\rho_d+p_d)=0 \label{conservation2}
\end{align}

with $\omega_m=p_m/\rho_m$ and $\omega_d=p_d/\rho_d$, the equation of state parameters for the two components.\\

As a particular choice, if we assume $p_d=-\rho_d=-\Lambda$, the cosmological constant, $\omega_m~(\text{i.e., $p_m$})=0$, then the above cosmological model is termed as $\Lambda$CDM model.  On the other hand, if the dark energy is introduced as a scalar field $\phi$ with self-interacting potential $V(\phi)$, we have the basic quintessence model i.e., 

\begin{align}
\rho_d=\rho_\phi=\frac{1}{2}\dot{\phi}^2+V(\phi),\label{equ5}
\end{align}
and
\begin{align}
    p_d=p_\phi=\frac{1}{2}\dot{\phi}^2-V(\phi)
\end{align}
then the conservation equation (\ref{conservation2}) becomes the Klein-Gordon equation
\begin{align}\label{KG}
\ddot{\phi}+3H\dot{\phi}+\frac{\partial V}{\partial \phi}=0
\end{align}
where $V'(\phi) \equiv \partial V/\partial \phi$.  This is usually known as the quintessence dark energy model.

On the other hand, the equation (\ref{FME2}) can be written as the following
\begin{align}\label{FME2a}
	2\dot{H}=-\dot{\phi}^2-(1+\omega_m)\rho_m.
\end{align}

As the above evolution equations are coupled and nonlinear in nature so it is hard to solve them analytically.  However, even without any solution, it is possible to get information about cosmological evaluation by using a dynamical system approach \cite{Oikonomou:2017ppp,Odintsov:2017tbc,Chakraborty:2021pkp, Chakraborty:2020vkp}.  This powerful method does not depend on the initial conditions or the evolution pattern at any intermediate instant \cite{wainwright_ellis_1997}.  Using suitable dimensionless variables the evolution equation can be converted into an autonomous set whose stable critical points correspond to different cosmological epochs.  Since late $90$'s, this approach has been widely used in various cosmological models to get inferences from critical points.  These studies are mainly confined to the background level and the critical points mostly correspond to behavior at early cosmological eras \cite{BAHAMONDE20181, JAWAD2022101127, CAPOZZIELLO2022137475, ACQUAVIVA2022101128}.  The dynamical analysis can
be enriched further with using bifurcation theory methods \cite{2022arXiv220206648B}. In this method, we inspect the phase transition of the evolution of the Universe. 
 We also characterize the generic and non-generic evolution of the Universe based on the initial and late time states of the evolution of the Universe \cite{Alho_2022, Humieja:2019ywy, Feng:2014fsa, Mishra:2021sza}.

	\section{Formation of the autonomous system and critical points determination\label{BES}}

The essence of the dynamical system approach is to transform the equations into an autonomous system using $\tau\equiv \ln a$ as the dynamical variable.	In the case of the quintessence scenario, the potential as of the form $V=V_0e^{-\lambda \phi}$ (exponential potential) \cite{Biswas:2016ces,article,article3,Singh:2019enu} is imposed on the scalar field. We introduce the auxiliary variables  
	\begin{align}
	x\equiv \frac{\dot{\phi}}{\sqrt{6}H}, \quad y \equiv \frac{\sqrt{V}}{\sqrt{3}H},\label{eqn1}
	\end{align}
where $x^2$  stands for the relative kinetic energy density of the scalar field and $y^2$ stands for its relative potential energy density of $\phi$. 
	
 The equations (\ref{KG}) and (\ref{FME2a}) can be expressed as the following autonomous system using the above auxiliary variables
		
	\begin{align}
		x'&=\frac{3}{2}x\left[2x^2+\gamma_{m}(1-x^2-y^2)\right]-3x+\sqrt{\frac{3}{2}} \lambda y^2,\label{eq1} \\ y'&=\frac{3}{2}y\left[2x^2+\gamma_{m}(1-x^2-y^2)\right]-\sqrt{\frac{3}{2}}\lambda xy ,\label{eq2}
	\end{align}
with $\gamma_m\equiv \omega_m+1$, in terms of which the various density parameters are expressed as $\Omega_d=x^2+y^2$, $\Omega_m=1-\Omega_d$, while $\omega_d=\frac{x^2-y^2}{x^2+y^2}$ and the expressions of total equation of state parameter $\omega_{total}$ and deceleration parameter $q$ are given by
$$
\begin{aligned}
	\omega_{total} &=2x^2-1+\gamma_m(1-x^2-y^2), \\
q&=\frac{3}{2}\gamma_m (1-x^2-y^2)+3x^2-1.
\end{aligned}
$$

In this paper, we study the discrete time dynamical systems associated with the system $(\ref{eq1}-\ref{eq2})$.   To study discrete time dynamical system analysis, first, we write the following proposition:\\

\begin{prop}
Let us consider the two dimensional nonlinear system of equations 
\begin{align}
\frac{dx}{dt}&=f(x,y),\label{eqq11} \\ \frac{dy}{dt}&=g(x,y).	\label{eqq12}
\end{align}
The discrete dynamical system corresponding to the above system is given by
\begin{align*}
x_{n+1}&=x_n+hf(x_n,y_n), \\
y_{n+1}&=y_n+hg(x_n,y_n).
\end{align*}

\end{prop}

\textsc{Proof}: We start with a discrete set of points $t_0,~t_1,\dots,~t_n,\dots$, with $h=t_{n+1}-t_n$ as step size.  Then, for $t_n \leq t <t_{n+1}$, we approximate $x(t)$ by $x(t_n)$ and $\frac{dx}{dt}=f(x(t),y(t))$ by $\frac{x(t_{n+1})-x(t_n)}{h}$ where $x(t)$ and $y(t)$ respectively denotes $\frac{\dot{\phi}(t)}{\sqrt{6}H(t)}$ and $\frac{\sqrt{V(t)}}{\sqrt{3}H(t)}$.  Equation (\ref{eqq11}) can be approximated as 
$$
x(t_{n+1})=x(t_n) + hf(x(t_n),y(t_n))
$$
Proceeding in similar way, from (\ref{eqq12}), we get
$$
y(t_{n+1})=y(t_n) + hg(x(t_n),y(t_n))
$$
Consideing $x(t_n)=x_n$, $x(t_{n+1})=x_{n+1}$, we can write the above two equations in the simpler form
\begin{align*}
	x_{n+1}&=x_n+hf(x_n,y_n), \\
	y_{n+1}&=y_n+hg(x_n,y_n).
\end{align*}

The discrete time dynamical systems associated with the system $(\ref{eq1}-\ref{eq2})$ can be expressed as 
\begin{align}
x_{n+1}&=h \left(\frac{3}{2}x_n\left[2x_n^2+\gamma_{m}(1-x_n^2-y_n^2)\right]-3x_n+\sqrt{\frac{3}{2}} \lambda y_n^2\right)+x_n,\label{eqn3} \\ y_{n+1}&=h\left(\frac{3}{2}y_n\left[2x_n^2+\gamma_{m}(1-x_n^2-y_n^2)\right]-\sqrt{\frac{3}{2}}\lambda x_n y_n\right)+y_n .\label{eqn4}
\end{align}

 Let us define the operator $W : \mathbb{R}^2\rightarrow \mathbb{R}^2$ by $W(x,y)=(\overline{x},\overline{y})$, where (considering the step size $h=1$ \cite{2022arXiv220206648B}) 
	\begin{align}
		\overline{x}&=\frac{3}{2}x\left[2x^2+\gamma_{m}(1-x^2-y^2)\right]-3x+\sqrt{\frac{3}{2}} \lambda y^2+x,\label{eq3} \\ \overline{y}&=\frac{3}{2}y\left[2x^2+\gamma_{m}(1-x^2-y^2)\right]-\sqrt{\frac{3}{2}}\lambda xy+y .\label{eq4}
  \end{align}
Here, we considered $x_{n+1}=\overline{x},y_{n+1}=\overline{y}$ and $x_n=x,y_n=y$ for simplicity of calculation.

\begin{definition}[Fixed points.]
	A point $z\in \mathbb{R}^2$ is called a fixed point \cite{2022arXiv220206648B} of $W$ if $W(z)=z$.
\end{definition}

By using the definition of fixed points, we can obtain the fixed points corresponding to the system $(\ref{eq3}-\ref{eq4})$ which are shown in Table \ref{T1}.  The existence of the fixed points is also shown in this table.

\begin{table}[h]
	\caption{\label{T1}Table shows the set of fixed points and their existence corresponding to the autonomous system $(\ref{eq3}-\ref{eq4})$:}
	\begin{tabular}{|c|c|c c|c|c|c|c|}
		
		\hline
		\begin{tabular}{@{}c@{}}$~~$\\ Fixed Points \\$~$\end{tabular}   & Existence& $x$ &$y$& $\Omega_d$ & $\omega_d$ & $\omega_{total}$& $q$\\ \hline\hline
		
		\begin{tabular}{@{}c@{}}$~~$\\$A$\\$~$\end{tabular}  & For all $\lambda$ &$0$ & $0$ & $0$ &NA &   $\gamma_m-1$ & $\frac{3}{2}\gamma_m-1$ \\ \hline 
		
		\begin{tabular}{@{}c@{}}$~~$\\$B$\\$~$\end{tabular}  & For all $\lambda$ &$1$ & $0$ &$1$ & $1$ &  $1$ & $2$ \\ \hline 
		
		\begin{tabular}{@{}c@{}}$~~$\\$C$\\$~$\end{tabular}  & For all $\lambda$ &$-1$ & $0$&$1$ & $1$&  $1$ & $2$   \\ \hline 
		
		\begin{tabular}{@{}c@{}}$~~$\\$D$\\$~$\end{tabular}  & $-\sqrt{6}\leq \lambda\leq \sqrt{6}$ &$\dfrac{\lambda}{\sqrt{6}}$ & $\sqrt{1-\dfrac{\lambda^2}{6}}$ & $1$ & $\frac{\lambda^2}{3}-1$ & $\frac{1}{3}(\lambda^2-3)$ & $\frac{\lambda^2}{2}-1$  \\ \hline 
		
		\begin{tabular}{@{}c@{}}$~~$\\$E$\\$~$\end{tabular}  & $\lambda^2>3\gamma_m$ & $\sqrt{\dfrac{3}{2}}\dfrac{\gamma_m}{\lambda}$ & $\sqrt{\dfrac{3(2-\gamma_m)\gamma_m}{2\lambda^2}}$ & $\frac{3\gamma_m}{\lambda^2}$ & $\gamma_m-1$&$\gamma_m-1$  & $\frac{3}{2}\gamma_m-1$  \\ \hline 
	\end{tabular}	
\end{table}

\bigbreak 

\section{STABILITY ANALYSIS}\label{III}

In this section, we will establish some simple but powerful criteria for the local stability of fixed points \cite{book:1378164,BISWAS2020145,Saha2018}. Fixed (equilibrium) points may be divided into two types: hyperbolic and nonhyperbolic.  A fixed point $x^*$ of a map $f$ is said to be hyperbolic if $|f'(x^*)|\neq 1$. Otherwise, it is nonhyperbolic. We will
	treat the stability of each type separately in Appendix \ref{A}. 
 To find the type of a fixed point of the system $(\ref{eq3}-\ref{eq4})$ we write the Jacobian matrix at $(x,y)$: 
$$
J(x,y)=\begin{pmatrix}
	\left(-2+\frac{3}{2}\gamma_m\right)+x^2 \left(9-\frac{9\gamma_m}{2}\right)-\frac{3}{2}\gamma_m y^2 & -3\gamma_m xy+\sqrt{6}\lambda y \\ \\
	6xy-3\gamma_m xy-\sqrt{\frac{3}{2}}\lambda y &  \left(\frac{3}{2}\gamma_m+1\right)-\sqrt{\frac{3}{2}}\lambda x+\left(3-\frac{3}{2}\gamma_m\right)x^2-\frac{9}{2}\gamma_m y^2
\end{pmatrix}
$$
Now we discuss the stability of all fixed points corresponding to the system $(\ref{eq3}-\ref{eq4})$.\\

$\bullet$ The fixed point $A$ exists for all $\lambda$.  For this solution, the kinetic term is subdominant compared to the matter contribution in equation (\ref{FME2a}).  The DE can represent either quitessence or any other exotic type fluid depending on the parameter $\gamma_m$.  Specially, for $0<\gamma_m<\frac{2}{3}$,  the scalar field behaves as quintessence like fluid and there exists an accelerated universe (since for this case, $-1<\omega_{total}<-\frac{1}{3}$, $q<0$) near the critical point whereas the scalar field DE behaves as cosmological constant for $\gamma_m=0$ and in this case the  cosmic evolution near the point characterizes the $\Lambda$CDM model ($\omega_{total}=0$, $q=-1$, $\Omega_d=0$) of evolution.

The eigenvalues of the Jacobian matrix $J(A)$ are $\lambda_1=-2+\frac{3}{2}\gamma_m$ and $\lambda_2=\frac{3}{2}\gamma_m +1$. The fixed point $A$ is hyperbolic if $\gamma_m\neq 0, \frac{2}{3}$ because for these values of $\gamma_m$ (that is, for $\gamma_m= 0, 2/3$), the absolute value of one of the eigenvalues of $J(A)$ is $1$ and the fixed point is nonhyperbolic if $\gamma_m=0$ or $\frac{2}{3}$.  Now we analyze the stability of the critical point $A$ for both of the cases (hyperbolic and nonhyperbolic).\\

\textbf{Hyperbolic case} $\left(\gamma_m\neq 0,\frac{2}{3}\right)$:\\

For the physically meaningful case: $0\leq \gamma_m< 2$.  So for the hyperbolic case we consider the value of $\gamma_m$ only when $\gamma_m\in \left(0,\frac{2}{3}\right)\cup\left(\frac{2}{3},2\right)$.  We can easily see that $0<\lambda_1<1$ for $\gamma_m\in \left(\frac{4}{3},2\right)$, $-1<\lambda_1<0$ for $\gamma_m\in \left(\frac{2}{3},\frac{4}{3}\right)$, $\lambda_1<-1$ if $\gamma_m \in \left(0,\frac{2}{3}\right)$ and $\lambda_2>1$ for $\gamma_m\in (0,2)$.  That means, for $\gamma_m\in \left(\frac{4}{3},2\right)$ we have $0<\lambda_1<1<\lambda_2$ which implies that the critical point $A$ is a \textit{saddle point}, for $\gamma_m\in \left(\frac{2}{3},\frac{4}{3}\right)$ we have $-1<\lambda_1<0, \lambda_2>1$ which suggests that the critical point $A$ is also a \textit{saddle point}, and for $\gamma_m\in \left(0,\frac{2}{3}\right)$ we have $\lambda_1<-1,\lambda_2>1$ which concludes that the critical point $A$ is a \textit{source}. \\

\textbf{Nonhyperbolic case} $\left(\gamma_m= 0~\text{or}~\frac{2}{3}\right)$:\\

To analyze the fixed point $A$ for this case, we use the center manifold theory.  For $\gamma_m=0$, the system $(\ref{eq3}-\ref{eq4})$ modifies to
$$
\begin{aligned}
	\overline{x}&=-2x+3x^3+\sqrt{\frac{3}{2}} \lambda y^2, \\
	\overline{y}&=y+3x^2 y-\sqrt{\frac{3}{2}}\lambda xy.
\end{aligned}
$$
Consider the map $F=\begin{pmatrix}
	f\\g
\end{pmatrix}$ defined by
$$
\begin{pmatrix}
	x\\\\y
\end{pmatrix}\longmapsto  \begin{pmatrix}
-2 & 0 \\ \\ 0 & 1
\end{pmatrix}\begin{pmatrix}
x \\ \\y
\end{pmatrix}+\begin{pmatrix}
3x^3+\sqrt{\frac{3}{2}}\lambda y^2 \\
3x^2y-\sqrt{\frac{3}{2}}\lambda  x y
\end{pmatrix}.
$$
Then,
$$
M_c=\left\{(x,y)\in \mathbb{R}^2 : x=h(y),h(0)=h'(0)=0\right\}.
$$

\begin{figure}
	\includegraphics[width=1\textwidth]{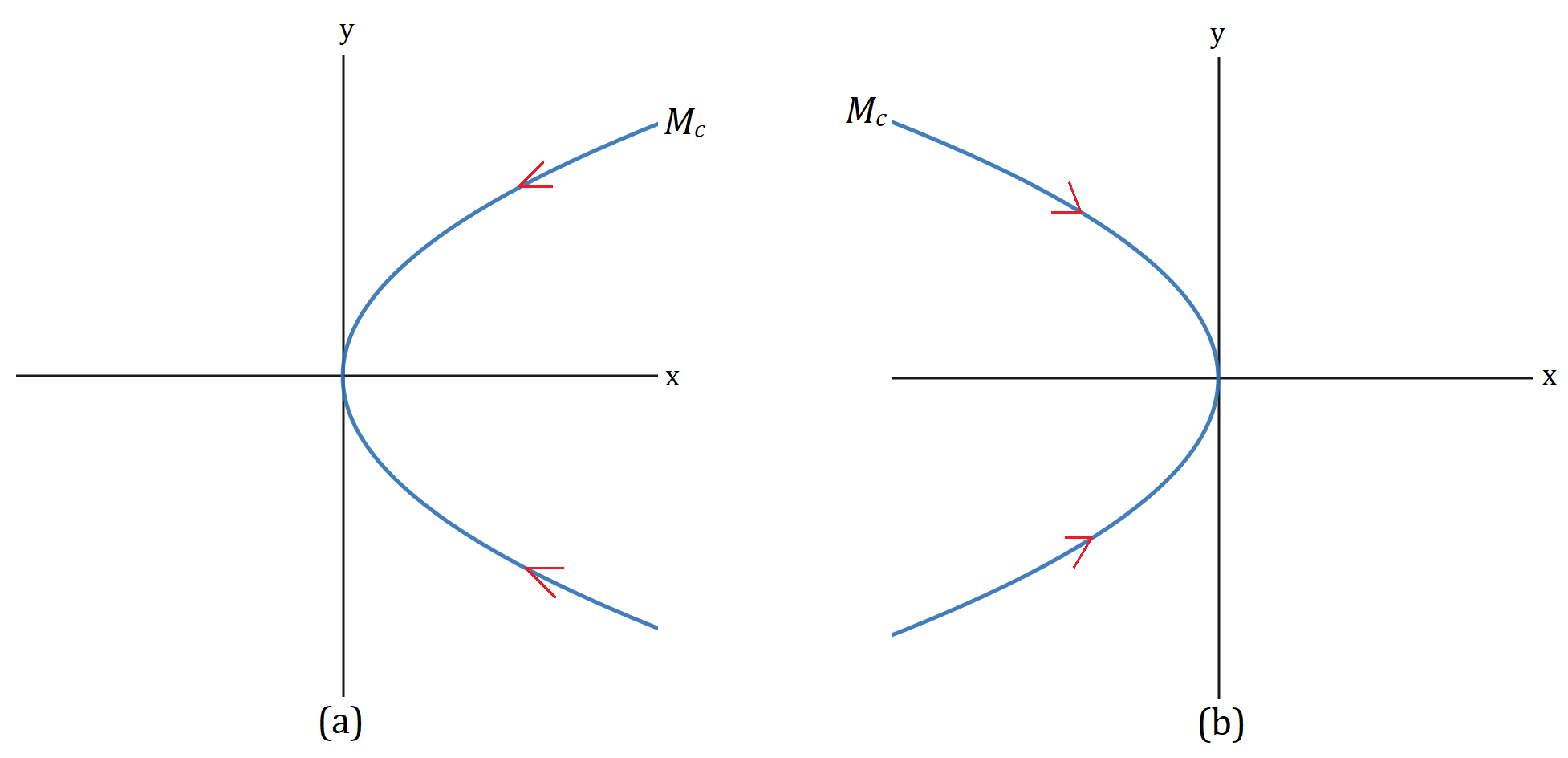}
	\caption{\label{Asymptotically_stable} The curve $h(y)=\frac{\lambda}{\sqrt{6}}y^2+\mathcal{O}(y^4)$ is the graph of the center manifold $M_c$.  The orbits on the $x$-axis oscillate but converge to the origin.  (a) is for $\lambda>0$ and (b) is for $\lambda<0$.}
\end{figure}

\begin{figure}
	\includegraphics[width=1\textwidth]{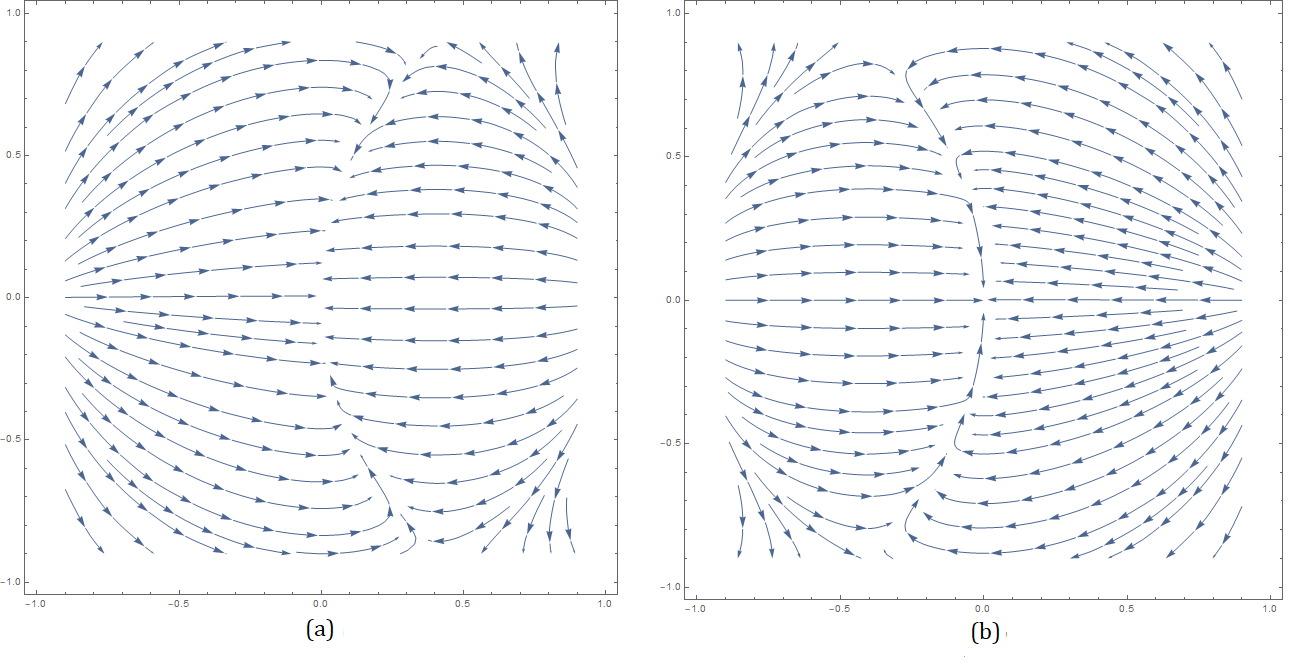}
	\caption{\label{c_1}The panel of the figures show the phase portrait of autonomous system $(\ref{eq1}-\ref{eq2})$ corresponding to the fixed point $A$. (a) is for $\lambda>0$ and (b) is for $\lambda<0$.  From these plots, we can easily observe that there exists a curve along which the system converges to the origin and that curve is nothing but the center manifold which we have obtained theoretically.}
\end{figure}

The function $h$ must satisfy Equation (\ref{cal})
$$
h\left(By+g(h(y),y)\right)-Ah(y)-f(h(y),y)=0
$$
or
\begin{align}
h\left(y+3yh^2(y)-\sqrt{\frac{3}{2}}\lambda y h(y)\right)+2h(y)-\left(3h^3(y)+\sqrt{\frac{3}{2}}\lambda y^2\right)=0.\label{eq9}
\end{align}
Let us assume that $h(y)$ takes the form
\begin{align}
	h(y)=c_1 y^2+c_2 y^3+\mathcal{O}(y^4).\label{eq10}
\end{align}
Then, substituting Equation (\ref{eq10}) in Equation (\ref{eq9}) and writing the terms upto degree $3$ yields
$$
c_1y^2+c_2 y^3+2(c_1y^2+c_2y^3)-\sqrt{\frac{3}{2}}\lambda y^2+\cdots =0.
$$
As we analyze the arbitrary neighborhood of the origin so by comparing both sides the coefficient of $y^2$ and $y^3$ yields
$$
\begin{aligned}
&	c_1+2c_1-\sqrt{\frac{3}{2}}\lambda=0\implies c_1=\frac{\lambda}{\sqrt{6}};\\
 &	c_2+2c_2=0 \implies c_2=0.
\end{aligned}
$$ 
Consequently, $h(y)=\frac{\lambda}{\sqrt{6}}y^2+\mathcal{O}(y^4)$ and the map $g$ on the center manifold is given by
$$
y \longmapsto y-\frac{\lambda^2}{2}y^3+\mathcal{O}(y^4).
$$
Notice that $y^*=0$ is a fixed point of $g$ at which $g'(0)=1$, $g''(0)=0$, and $g'''(0)=-3\lambda^2<0$.  This implies by Theorem \ref{2} that the critical point $A$ is \textit{asymptotically stable} (see figure \ref{Asymptotically_stable}) under the map $F=\begin{pmatrix}
	f\\g
\end{pmatrix}$.  To verify our conclusion, we also draw the phase portrait of the original system numerically (see figure \ref{c_1}).  From that plot, we can conclude that our theoretical result is exactly the same as the result which is obtained by plotting the system numerically \cite{Oikonomou:2017ppp,Odintsov:2017tbc}.\par

For $\gamma_m=\frac{2}{3}$, the calculation of the center manifold is shown in Appendix \ref{B}.  From the calculation and the second equation corresponding to the autonomous system, we can conclude that the critical point $A$ is unstable in nature for this case.\\


$\bullet$ The fixed point $B$ corresponds to solution where the constraint equation (\ref{FME2a}) is dominated by the kinetic energy of the scalar field with a stiff equation of state $\omega_d=1$ and there exist a decelerating phase of the universe.

The eigenvalues of the Jacobian matrix $J(B)$ are $\lambda_1=7-3\gamma_m$ and $\lambda_2=4-\sqrt{\frac{3}{2}}\lambda$. The fixed point $B$ is hyperbolic if $\lambda\neq \sqrt{6}, \frac{5\sqrt{2}}{\sqrt{3}}$ because for these values of $\lambda$ (that is, for $\lambda=\sqrt{6}, \frac{5\sqrt{2}}{\sqrt{3}}$), the absolute value of one of the eigenvalues of $J(B)$ is $1$ and the fixed point is nonhyperbolic if $\lambda=\sqrt{6}$ or $\lambda=\frac{5\sqrt{2}}{\sqrt{3}}$.  Now we analyze the stability of the critical point $B$ for both of the cases (hyperbolic and nonhyperbolic).\\

\textbf{Hyperbolic case} $\left(\lambda\neq \sqrt{6}, \frac{5\sqrt{2}}{\sqrt{3}}\right)$:\\

For the physically meaningful case: $0\leq \gamma_m< 2$.  We can easily see that $\lambda_1>1$ for all $\gamma_m\in \left[0,2\right)$.  Further, we can see that $\lambda_2>1$ when $\lambda<\sqrt{6}$, $0<\lambda_2<1$ when $\sqrt{6}<\lambda<\frac{4\sqrt{2}}{\sqrt{3}}$, $-1<\lambda_2<0$ when $\frac{4\sqrt{2}}{\sqrt{3}}<\lambda<\frac{5\sqrt{2}}{\sqrt{3}}$, and $\lambda_2<-1$ when $\lambda>\frac{5\sqrt{2}}{\sqrt{3}}$.  It follows that the fixed point $B$ is a \textit{source} for $\lambda<\sqrt{6}$ or $\lambda>\frac{5\sqrt{2}}{\sqrt{3}}$, and \textit{saddle point} for $\lambda\in\left(\sqrt{6},\frac{4\sqrt{2}}{\sqrt{3}}\right)\cup\left(\frac{4\sqrt{2}}{\sqrt{3}},\frac{5\sqrt{2}}{\sqrt{3}}\right)$.\\

\textbf{Nonhyperbolic case} $\left(\text{$\lambda=\sqrt{6}$ or $\lambda=\frac{5\sqrt{2}}{\sqrt{3}}$}\right)$:\\

To apply center manifold theory, first, we have to transform the critical point $B$ to the origin.  To do that, we consider the shifting transformation $x=X+1$, $y=Y$ and after applying the shifting transformation and putting $\lambda=\sqrt{6}$, we can determine that
the equations of center manifold \cite{book:1378164} is given by $h(Y)=-\frac{1}{2}Y^2+\mathcal{O}(Y^4)$ and the map $g$ on the center manifold is 
$$
Y \longmapsto Y-\frac{3}{2}Y^3+\mathcal{O}(Y^5).
$$
Notice that $Y^*=0$ is a fixed point of $g$ at which $g'(0)=1$, $g''(0)=0$, and $g'''(0)=-9<0$.  This implies by Theorem \ref{2} that the flow is stable along the center manifold but the critical point is \textit{unstable (semistable)} as $\lambda_1>1$ for all $\gamma_m$.\\

After applying the shifting transformation and putting $\lambda=\frac{5\sqrt{2}}{\sqrt{3}}$, the equation of the center manifold is given by 
 $h(Y)=\frac{10-3\gamma_m}{2(3\gamma_m -6)}Y^2+\mathcal{O}(Y^4)$ and the map $g$ on the center manifold is given by
$$
Y \longmapsto -Y+\frac{5}{6}\left(\frac{2-3\gamma_m}{\gamma_m-2}\right)Y^3+\mathcal{O}(Y^5).
$$
Notice that $Y^*=0$ is a fixed point of $g$ at which $g'(0)=-1$.  Now we determine the Schwarzian derivative, $Sg$, of the function $g$.
$$
Sg(0)=-g'''(0)-\frac{3}{2}\left[g''(0)\right]^2=\frac{5(3\gamma_m-2)}{\gamma_m-2}.
$$
We can easily check that $Sg(0)>0$ while $\gamma_m<\frac{2}{3}$ and $Sg(0)<0$ while $\frac{2}{3}<\gamma_m<2$.  Thus by using Theorem \ref{3}, we can conclude that $Y^*=0$ is stable along the center manifold when $\frac{2}{3}<\gamma_m<2$ but as $\lambda_1>1$ for all $\gamma_m$, this implies that $Y^*=0$ is semi stable for $\frac{2}{3}<\gamma_m<2$ and unstable under the map $g$ when $\gamma_m<\frac{2}{3}$.  This implies by Theorem \ref{7} that for $\lambda=\frac{5\sqrt{2}}{\sqrt{3}}$ the fixed point $B$ is \textit{semi stable} under the map $F$ for $\frac{2}{3}<\gamma_m<2$ and \textit{unstable} under the map $F$ for $\gamma_m<\frac{2}{3}$.\\

$\bullet$ Similar to the fixed point $B$, the fixed point $C$ also corresponds to solution where the constraint equation (\ref{FME2a}) is dominated by the kinetic energy of the scalar field with a stiff equation of state $\omega_d=1$ and there exist a decelerating phase of the universe.

The eigenvalues of the Jacobian matrix $J(C)$ are $\lambda_1=7-3\gamma_m$ and $\lambda_2=4+\sqrt{\frac{3}{2}}\lambda$. The fixed point $C$ is hyperbolic if $\lambda\neq -\sqrt{6},~ -\frac{5\sqrt{2}}{\sqrt{3}}$ because for these two values of $\lambda$, the absolute value of one of the eigenvalues of $J(C)$ is $1$ and the fixed point is nonhyperbolic if $\lambda=-\sqrt{6}$ or $\lambda=-\frac{5\sqrt{2}}{\sqrt{3}}$.  Now we analyze the stability of the critical point $C$ for both of the cases (hyperbolic and nonhyperbolic).\\

\textbf{Hyperbolic case} $\left(\lambda\neq -\sqrt{6}, -\frac{5\sqrt{2}}{\sqrt{3}}\right)$:\\

For the physically meaningful case: $0\leq \gamma_m<2$.  We can easily see that $\lambda_1>1$ for all $\gamma_m\in \left[0,2\right)$.  Further, we can see that $\lambda_2>1$ when $\lambda>-\sqrt{6}$, $0<\lambda_2<1$ when $-\frac{4\sqrt{2}}{\sqrt{3}}<\lambda<-\sqrt{6}$, $-1<\lambda_2<0$ when $-\frac{5\sqrt{2}}{\sqrt{3}}<\lambda<-\frac{4\sqrt{2}}{\sqrt{3}}$, and $\lambda_2<-1$ when $\lambda<-\frac{5\sqrt{2}}{\sqrt{3}}$.  It follows that the fixed point $C$ is a \textit{source} for $\lambda\in(-\infty,-\frac{5\sqrt{2}}{\sqrt{3}})\cup(-\sqrt{6},\infty)$, and  \textit{saddle point} for $\lambda\in\left(-\frac{4\sqrt{2}}{\sqrt{3}},-\sqrt{6}\right)\cup\left(-\frac{5\sqrt{2}}{\sqrt{3}},-\frac{4\sqrt{2}}{\sqrt{3}}\right)$.\\

\textbf{Nonhyperbolic case} $\left(\text{$\lambda=-\sqrt{6}$ or $\lambda=-\frac{5\sqrt{2}}{\sqrt{3}}$}\right)$:\\

To apply center manifold theory, first, we have to transform the critical point $B$ to the origin.  To do that, we consider the shifting transformation $x=X-1$, $y=Y$ and after applying the shifting transformation and putting $\lambda=-\sqrt{6}$, the equation of the center manifold can be written as 
 $h(Y)=\frac{1}{2}Y^2+\mathcal{O}(Y^4)$ and the map $g$ on the center manifold is given by
$$
Y \longmapsto Y-\frac{3}{2}Y^3+\mathcal{O}(Y^5).
$$
Notice that $Y^*=0$ is a fixed point of $g$ at which $g'(0)=1$, $g''(0)=0$, and $g'''(0)=-9<0$.  This implies by Theorem \ref{2} that the critical point $C$ is stable along the center manifold but as $\lambda_1>1$ for all $\gamma_m$ so the critical point is \textit{semi stable} in nature.\\

After applying the shifting transformation and putting $\lambda=-\frac{5\sqrt{2}}{\sqrt{3}}$,
the equation of the center manifold can be written as $h(Y)=\frac{3\gamma_m-10}{2(3\gamma_m -6)}Y^2+\mathcal{O}(Y^4)$ and the map $g$ on the center manifold is given by
$$
Y \longmapsto -Y+\frac{5}{6}\left(\frac{2-3\gamma_m}{\gamma_m-2}\right)Y^3+\mathcal{O}(Y^5).
$$
Notice that $Y^*=0$ is a fixed point of $g$ at which $g'(0)=-1$.  Now we determine the Schwarzian derivative, $Sg$, of the function $g$.
$$
Sg(0)=-g'''(0)-\frac{3}{2}\left[g''(0)\right]^2=\frac{5(3\gamma_m-2)}{\gamma_m-2}.
$$
We can easily check that $Sg(0)>0$ while $\gamma_m<\frac{2}{3}$ and $Sg(0)<0$ while $\frac{2}{3}<\gamma_m<2$.  Thus by using Theorem \ref{3}, we conclude that $Y^*=0$ is asymptotically stable for $\frac{2}{3}<\gamma_m<2$ and unstable under the map $g$ when $\gamma_m<\frac{2}{3}$.  This implies by Theorem \ref{7} that for $\lambda=-\frac{5\sqrt{2}}{\sqrt{3}}$, the fixed point $C$ is \textit{asymptotically stable} under the map $F$ for $\frac{2}{3}<\gamma_m<2$ and \textit{unstable} under the map $F$ for $\gamma_m<\frac{2}{3}$.\\

$\bullet$ The fixed point $D$ exists for $-\sqrt{6}\leq \lambda \leq \sqrt{6}$.  Varying $\lambda$ we get a non-isolated set of critical points which are completely dominated by kinetic energy ($\Omega_d=1$ and $\dot{\phi}\neq 0$).  It is to be noted that the DE can represent either quintessence or any other exotic type fluid depending on the parameter $\lambda$.  The restriction $0<\lambda^2<2$  implies that the
scalar field behaves as quintessence like fluid and there exists an accelerated universe (since for this case, $-1<\omega_{total}<-1/3$, $q<0$) near these critical points whereas the scalar field DE behaves as dust for $\lambda^2=3$ ($\omega_{\phi}=0$) and in this case, the solutions insinuate decelerating phase of the cosmic evolution.


The eigenvalues of the Jacobian matrix $J(D)$ are $\lambda_1=\frac{1}{2}(\lambda^2-4)$ and $\lambda_2=1+\lambda^2-3\gamma_m$.  The fixed point $D$ is hyperbolic if $\gamma_m\neq \frac{\lambda^2}{3},\frac{\lambda^2+2}{3}$; $\lambda\neq \pm \sqrt{6},~ \pm \sqrt{2}$ because for these values of $\gamma_m$ and $\lambda$, the absolute value of one of the eigenvalues of $J(D)$ is $1$ and the fixed point is nonhyperbolic if $\gamma_m=\frac{\lambda^2}{3}$ or $\gamma_m=\frac{\lambda^2+2}{3}$ or $\lambda=\pm \sqrt{6}$ or $\lambda=\pm \sqrt{2}$.  Now we analyze the stability of the critical point $D$ for both of the cases (hyperbolic and nonhyperbolic).\\

\textbf{Hyperbolic case} $\left(\gamma_m\neq \frac{\lambda^2}{3},\frac{\lambda^2+2}{3}; \lambda\neq \pm \sqrt{6},~ \pm \sqrt{2}\right)$:\\

For $\gamma_m\in [0,2)$, we can see that $\lambda_1,\lambda_2<1$ while $\lambda^2<3\gamma_m$; $\lambda_1<1, \lambda_2>1$ while $3\gamma_m<\lambda^2<6$; $\lambda_1,\lambda_2>1$ while $\lambda^2>6$.  It follows that for $\gamma_m\in [0,2)$, the fixed point $D$ is a \textit{stable node (sink)} while $\lambda^2<3\gamma_m$, saddle point while $3\gamma_m<\lambda^2<6$ and \textit{unstable node (source)} while $\lambda^2>6$. \\

\textbf{Nonhyperbolic case} $\left(\gamma_m= \frac{\lambda^2}{3}~\text{or}~\gamma_m=\frac{\lambda^2+2}{3}~\text{or}~ \lambda= \pm \sqrt{6},~\text{or}~\lambda= \pm \sqrt{2}\right)$:\\

First of all, note that for $\lambda=\sqrt{6}$ the coordinate of critical point $D$ is equivalent to the coordinate of critical point $B$ and the stability of the critical point $B$ for $\lambda=\sqrt{6}$ already discussed.  Further note that for $\lambda=-\sqrt{6}$ the coordinate of critical point $D$ is equivalent to the coordinate of critical point $C$ and the stability of the critical point $C$ for $\lambda=-\sqrt{6}$ already discussed.  So here we will not discuss the stability of the fixed point $D$ for $\lambda=\pm\sqrt{6}$.  Now we discuss the stability of the fixed point $D$ for $\lambda=\sqrt{2}$.\\

To analyze the fixed point $D$ for $\lambda=\sqrt{2}$, we use center manifold theory.  To apply center manifold theory, first, we have to transform the critical point $D$ to the origin.  To do that, we consider the shifting transformation $x=X+\frac{\lambda}{\sqrt{6}}$, $y=Y+\sqrt{1-\frac{\lambda^2}{6}}$ and after applying the shifting transformation and putting $\lambda=\sqrt{2}$, the discrete dynamical system associated with the system $(\ref{eq1}-\ref{eq2})$can be written as 

$$
\begin{aligned}
\overline{X}&=X-\frac{3}{2}  X^3 \gamma _m-\frac{3\sqrt{3}}{2}  X^2 \gamma _m-X \gamma _m-\frac{3}{2} X Y^2 \gamma _m-\sqrt{6} X Y \gamma _m-\frac{\sqrt{3}}{2}  Y^2 \gamma _m-\sqrt{2} Y \gamma _m+3 X^3+3 \sqrt{3} X^2\\&\phantom{=}+\sqrt{3} Y^2+2 \sqrt{2} Y, \\
\overline{Y} &=Y-\sqrt{\frac{3}{2}} X^2 \gamma _m-\frac{3}{2} X^2 Y \gamma _m-\sqrt{2} X \gamma _m-\sqrt{3} X Y \gamma _m-\frac{3}{2}  Y^3 \gamma _m-3 \sqrt{\frac{3}{2}} Y^2 \gamma _m-2 Y \gamma _m+3 X^2 Y+\sqrt{6} X^2\\&\phantom{=}+\sqrt{3} X Y+\sqrt{2} X .
\end{aligned}
$$

Next, we use a coordinate transformation $\mathbf{u}=P\mathbf{v}$ where $\mathbf{u}=(X~ Y)^T$, $\mathbf{v}=(u~ v)^T$, $P=(\mathbf{v_1}~\mathbf{v_2})$, and $\mathbf{v_1}$, $\mathbf{v_2}$ are the eigenvectors corresponding to the eigenvalues $\lambda_1$, $\lambda_2$ respectively.  The equation of center manifold can be written as $h(u)=\frac{9\sqrt{3}(\gamma_m-2)}{2(2-3\gamma_m)(4-3\gamma_m)}u^2+\mathcal{O}(u^3)$ and the map $f$ on the center manifold is given by
$$
u \longmapsto -u+\frac{1}{4-3\gamma_m}\left(-5\sqrt{6}+9\sqrt{\frac{3}{2}}\gamma_m\right)u^2+\frac{3 \left(27 \gamma _m^3-54 \gamma _m^2-12 \gamma _m+56\right)}{2 \left(4-3 \gamma _m\right){}^2 \left(3 \gamma _m-2\right)}u^3+\mathcal{O}(u^4).
$$
Notice that $u^*=0$ is a fixed point of $f$ at which $f'(0)=-1$.  Now we determine the Schwarzian derivative, $Sf$, of the function $f$.
$$
Sf(0)=-f'''(0)-\frac{3}{2}\left[f''(0)\right]^2=-\frac{162 \left(15\gamma_m^3-42\gamma_m^2+36\gamma_m-8\right)}{(4-3\gamma_m)^2(3\gamma_m-2)}.
$$
We can check that $Sf(0)<0$ when $\gamma_m<0.342$ or $\gamma_m>\frac{2}{3}$ and while $0.342<\gamma_m<\frac{2}{3}$ then $Sf(0)>0$.  Thus by using Theorem \ref{3}, we can conclude that $u^*=0$ is \textit{asymptotically stable} under the map $f$ when $\gamma_m<0.342$ or $\gamma_m>\frac{2}{3}$ and \textit{unstable} under the map $f$ while $0.342<\gamma_m<\frac{2}{3}$.  This implies by Theorem \ref{7} that for $\lambda=\sqrt{2}$ the fixed point $D$ is \textit{asymptotically stable} under the map $F$ when $\gamma_m<0.342$ or $\gamma_m>\frac{2}{3}$ and \textit{unstable} under the map $F$ when $0.342<\gamma_m<\frac{2}{3}$.\\

For $\lambda=-\sqrt{2}$, after considering the shifting transformation $x=X+\frac{\lambda}{\sqrt{6}}$, $y=Y+\sqrt{1-\frac{\lambda^2}{6}}$ and putting $\lambda=-\sqrt{2}$, the system $(\ref{eq3}-\ref{eq4})$ modifies to
$$
\begin{aligned}
\overline{X}&=X-\frac{1}{2} 3 X^3 \gamma _m+\frac{3}{2} \sqrt{3} X^2 \gamma _m-X \gamma _m-\frac{3}{2} X Y^2 \gamma _m-\sqrt{6} X Y \gamma _m+\frac{1}{2} \sqrt{3} Y^2 \gamma _m+\sqrt{2} Y \gamma _m+3 X^3\\&\phantom{=}-3 \sqrt{3} X^2-\sqrt{3} Y^2-2 \sqrt{2} Y, \\
\overline{Y} &=Y-\sqrt{\frac{3}{2}} X^2 \gamma _m-\frac{3}{2} X^2 Y \gamma _m+\sqrt{2} X \gamma _m+\sqrt{3} X Y \gamma _m-\frac{1}{2} 3 Y^3 \gamma _m-3 \sqrt{\frac{3}{2}} Y^2 \gamma _m-2 Y \gamma _m+3 X^2 Y\\&\phantom{=}+\sqrt{6} X^2-\sqrt{3} X Y-\sqrt{2} X .
\end{aligned}
$$
Next, we use a coordinate transformation $\mathbf{u}=P\mathbf{v}$ where $\mathbf{u}=(X~ Y)^T$, $\mathbf{v}=(u~ v)^T$, $P=(\mathbf{v_1}~\mathbf{v_2})$, and $\mathbf{v_1}$, $\mathbf{v_2}$ are the eigenvectors corresponding to the eigenvalues $\lambda_1$, $\lambda_2$ respectively.  The equation of center manifold can be written as $h(u)=\frac{9\sqrt{3}(2-\gamma_m)}{2(2-3\gamma_m)(4-3\gamma_m)}u^2+\mathcal{O}(u^3)$ and the map $f$ on the center manifold is given by
$$
u \longmapsto -u+\frac{1}{4-3\gamma_m}\left(-5\sqrt{6}+9\sqrt{\frac{3}{2}}\gamma_m\right)u^2+\frac{3 \left(27 \gamma_m^3-54 \gamma_m^2-12 \gamma_m+56\right)}{2 \left(9 \gamma_m^2-18 \gamma_m+8\right)}u^3+\mathcal{O}(u^4).
$$
Notice that $u^*=0$ is a fixed point of $f$ at which $f'(0)=-1$.  Now we determine the Schwarzian derivative, $Sf$, of the function $f$.
$$
Sf(0)=-f'''(0)-\frac{3}{2}\left[f''(0)\right]^2=-\frac{9 \left(81 \gamma_m^4-27 \gamma_m^3-522 \gamma_m^2+876 \gamma_m-424\right)}{(4-3 \gamma_m)^2 (3 \gamma_m-2)}.
$$
We can check that $Sf(0)<0$ while $\gamma_m>1.245$ or $\gamma_m<\frac{2}{3}$ and $Sf(0)>0$ while $\frac{2}{3}<\gamma_m<1.245$.  Thus by using Theorem \ref{3}, we can conclude that $u^*=0$ is asymptotically stable under the map $f$ when $\gamma_m>1.245$ or $\gamma_m<\frac{2}{3}$ and unstable under the map $g$ when $\frac{2}{3}<\gamma_m<1.245$.  This implies by Theorem \ref{7} that for $\lambda=-\sqrt{2}$ the fixed point $D$ is \textit{asymptotically stable} under the map $F$ when $\gamma_m>1.245$ or $\gamma_m<\frac{2}{3}$ and \textit{unstable} under the map $F$ when $\frac{2}{3}<\gamma_m<1.245$.\\

To analyze the fixed point $D$ for $\gamma_m=\frac{\lambda^2}{3}$, we use center manifold theory.  To apply center manifold theory, first, we have to transform the critical point $D$ to the origin.  To do that, we consider the shifting transformation $x=X+\frac{\lambda}{\sqrt{6}}$, $y=Y+\sqrt{1-\frac{\lambda^2}{6}}$ and after applying the shifting transformation and putting $\gamma_m=\frac{\lambda^2}{3}$, the system $(\ref{eq3}-\ref{eq4})$ modifies to
$$
\begin{pmatrix}
	X \\\\ Y
\end{pmatrix}\longmapsto  \begin{pmatrix}
-2+\frac{3}{2}\lambda^2-\frac{\lambda^4}{6} & \sqrt{6}\left(1-\frac{\lambda^2}{6}\right)^{3/2}\\ \\ \sqrt{\frac{3}{2}}\lambda \sqrt{1-\frac{\lambda^2}{6}}\left(1-\frac{\lambda^2}{3}\right) & 1-\lambda^2+\frac{\lambda^4}{6}
\end{pmatrix}\begin{pmatrix}
	X \\ \\ Y
\end{pmatrix}+\begin{pmatrix}
	non\\ linear \\terms.
\end{pmatrix}.
$$

Next, we use a coordinate transformation $\mathbf{u}=P\mathbf{v}$ where $\mathbf{u}=(X~ Y)^T$, $\mathbf{v}=(u~ v)^T$, $P=(\mathbf{v_1}~\mathbf{v_2})$, and $\mathbf{v_1}$, $\mathbf{v_2}$ are the eigenvectors corresponding to the eigenvalues $\lambda_1$, $\lambda_2$ respectively. The equation of center manifold can be represented as $h(u)=-\sqrt{\frac{3}{2}}\frac{\lambda^2(2\lambda^2-3)}{\sqrt{6-\lambda^2}(\lambda^2-3)^2}u^2+\mathcal{O}(u^3)$ and the map $f$ on the center manifold is given by
$$
u \longmapsto u+\frac{\sqrt{6}\lambda^2(6-\lambda^2)}{(\lambda^2-3)^2}u^2+\frac{\lambda^2\sqrt{6-\lambda^2}(9-12\lambda^2)}{(\lambda^2-3)^3}u^3+\mathcal{O}(u^4).
$$

Notice that $u^*=0$ is a fixed point of $f$ at which 
$$
\begin{aligned}
	f'(0)&=1, \\ f''(0)&=\frac{2\sqrt{6}\lambda^2(6-\lambda^2)}{(\lambda^2-3)^2}, \\ f'''(0)&=\frac{18\lambda^2(3-4\lambda^2)\sqrt{6-\lambda^2}}{(\lambda^2-3)^3}.
\end{aligned}
$$
 This implies by Theorem \ref{2} that the critical point $D$ is \textit{unstable (semistable)} when $\lambda\neq0,\pm\sqrt{6}$.  For $\lambda=0$ or $\lambda=\pm\sqrt{6}$ no conclusion can be determined.  Also note that for $\lambda=\pm\sqrt{6}$, the corresponding value of $\gamma_m$ is $\gamma_m=2$ which is not also physically meaningful case.\\

To analyze the fixed point $D$ for $\gamma_m=\frac{\lambda^2+2}{3}$, we use center manifold theory.  To apply center manifold theory, first, we have to transform the critical point $D$ to the origin.  To do that, we consider the shifting transformation $x=X+\frac{\lambda}{\sqrt{6}}$, $y=Y+\sqrt{1-\frac{\lambda^2}{6}}$ and after applying the shifting transformation and putting $\gamma_m=\frac{\lambda^2+2}{3}$, the system $(\ref{eq3}-\ref{eq4})$ modifies to
$$
\begin{pmatrix}
	X \\\\ Y
\end{pmatrix}\longmapsto  \begin{pmatrix}
-\frac{\lambda ^4}{6}+\frac{7 \lambda ^2}{6}-2 & \frac{2}{3} \lambda  \sqrt{6-\lambda ^2}-\frac{1}{6} \lambda ^3 \sqrt{6-\lambda ^2} \\ \\ \frac{1}{6} \lambda  \sqrt{6-\lambda ^2}-\frac{1}{6} \lambda ^3 \sqrt{6-\lambda ^2} & \frac{\lambda ^4}{6}-\frac{2 \lambda ^2}{3}-1
\end{pmatrix}\begin{pmatrix}
	X \\ \\ Y
\end{pmatrix}+\begin{pmatrix}
	non\\ linear \\terms.
\end{pmatrix}.
$$
Next, we use a coordinate transformation $\mathbf{u}=P\mathbf{v}$ where $\mathbf{u}=(X~ Y)^T$, $\mathbf{v}=(u~ v)^T$, $P=(\mathbf{v_1}~\mathbf{v_2})$, and $\mathbf{v_1}$, $\mathbf{v_2}$ are the eigenvectors corresponding to the eigenvalues $\lambda_1$, $\lambda_2$ respectively.  The equation of centermanifold can be represented as $h(u)=\sqrt{\frac{3}{2}}\frac{\lambda}{\sqrt{6-\lambda^2}} \left(\frac{2\lambda^6-15\lambda^4+30\lambda^2-8}{2-\lambda^2}\right)u^2+\mathcal{O}(u^3)$ and the map $f$ on the center manifold is given by
$$
\begin{aligned}
u & \longmapsto -u+\frac{-6 \sqrt{6} \left(\lambda ^2-6\right) \left(\lambda ^8-12 \lambda ^6+41 \lambda ^4-48 \lambda ^2+12\right) }{2(\lambda^2-2)}u^2\\&\phantom{\longmapsto}+\frac{9 \sqrt{6-\lambda ^2} \left(2 \lambda ^{12}-47 \lambda ^{10}+357 \lambda ^8-1168 \lambda ^6+1692 \lambda ^4-896 \lambda ^2+96\right) }{2(\lambda^2-2)}u^3+\mathcal{O}(u^4).
\end{aligned}
$$

Notice that $u^*=0$ is a fixed point of $f$ at which $f'(0)=-1$.  Now we determine the Schwarzian derivative, $Sf$, of the function $f$.
$$
\begin{aligned}
Sf(0)&=-f'''(0)-\frac{3}{2}\left[f''(0)\right]^2\\&=-6 \left(\frac{6 \sqrt{6} \left(\lambda ^2-6\right) \left(\lambda ^8-12 \lambda ^6+41 \lambda ^4-48 \lambda ^2+12\right)}{2 \left(\lambda ^2-2\right)}\right)^2\\  & \phantom{=} -\frac{27 \sqrt{6-\lambda ^2} \left(2 \lambda ^{12}-47 \lambda ^{10}+357 \lambda ^8-1168 \lambda ^6+1692 \lambda ^4-896 \lambda ^2+96\right)}{\lambda ^2-2}.
\end{aligned}
$$

By considering various values of $\lambda$, we can check that $Sf(0)<0$.  As $-\sqrt{6}<\lambda<\sqrt{6}$, to determine the value of $Sf(0)$, we have taken $\lambda=0,\pm 1, \pm 2$ and see that for these values of $\lambda$, $Sf(0)<0$.  Thus by using Theorem \ref{3}, we can conclude that $u^*=0$ is asymptotically stable under the map $f$.  This implies by Theorem \ref{7} that for $\gamma_m=\frac{\lambda^2+2}{3}$ the fixed point $D$ is \textit{asymptotically stable} under the map $F$ for all values of $\lambda\in (-\sqrt{6},\sqrt{6})$.\\

$\bullet$ The fixed point $E$ exists for $\lambda^2>3\gamma_{m}$.  Scaling solutions are  represented by these critical points. These non-isolated set of critical points are hyperbolic or normally-hyperbolic in nature  depending on the values of $\lambda$ and $\gamma_{m}$.  At $\gamma_{m}=\frac{2}{3}$, the scalar field DE behaves as a quintessence boundary.  The cosmic coincidence problem can be alleviated by these points depending on the values of $\lambda$ and $\gamma_{m}$.

\begin{figure}[!]
	\includegraphics[width=0.8\textwidth]{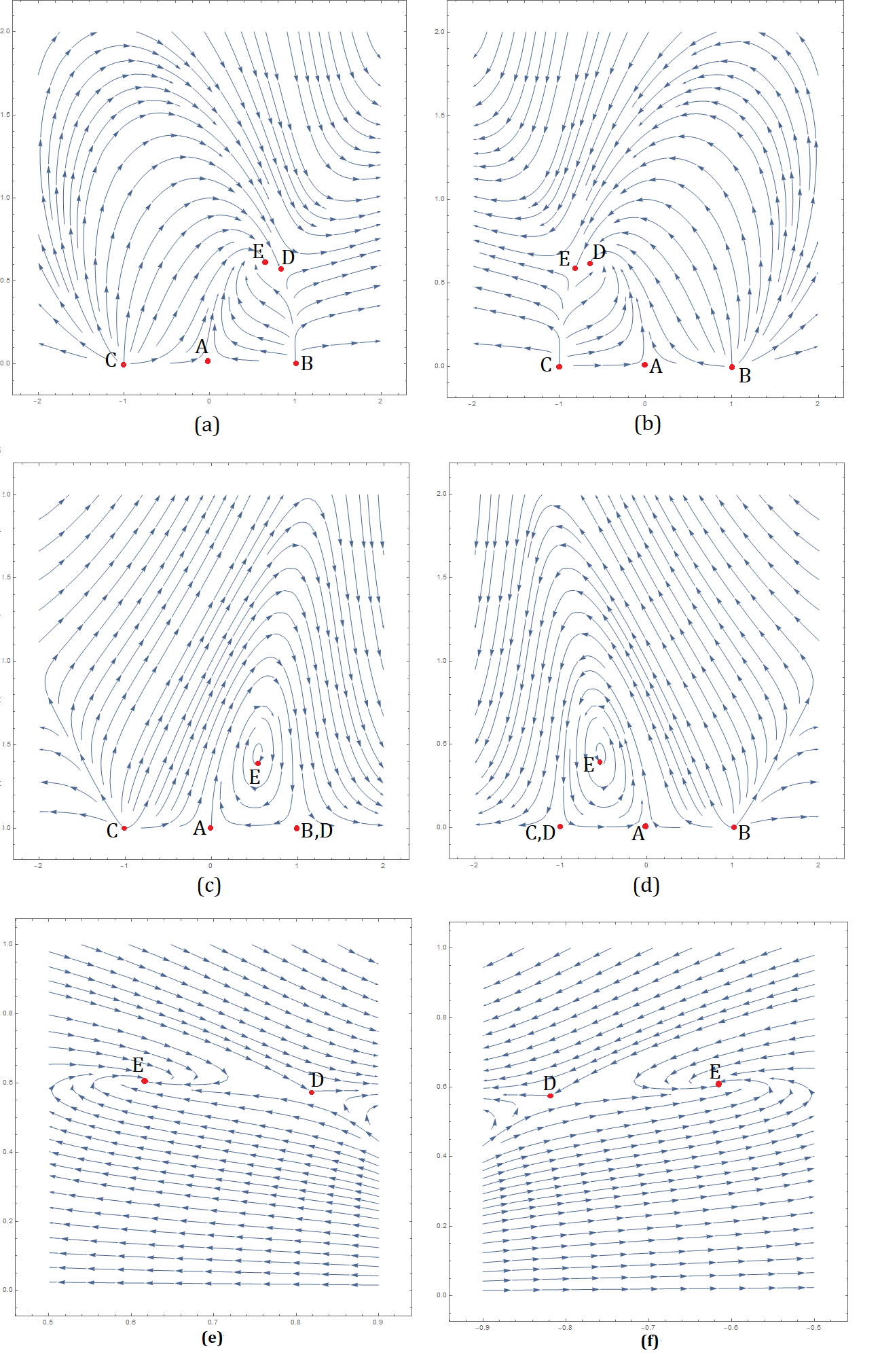}
	\caption{\label{E_1}  These figures show the behavior of the trajectories near each of the critical points.  Especially in this case, we want to focus on the phase portrait near the critical point $E$.  (a) is for $\lambda=2,\gamma_m=1$, (b) is for $\lambda=-2,\gamma_m=1$, (c) is for $\lambda=\sqrt{6},\gamma_m=\frac{4}{3}$, and (d) is for $\lambda=-\sqrt{6},\gamma_m=\frac{4}{3}$.  Figures (e) and (f) show the phase portraits in a very small neighborhood of the critical points $D$ and $E$ to distinguish the behavior of stability of the critical points for $\lambda=\pm2$, $\gamma_m=1$.}
\end{figure}

The critical point $E$ has more complicated stability conditions. The stability status of $E$ is depicted in Figure \ref{E_1}. In those figures, we have shown the phase portrait near the critical point $E$ numerically for several physically meaningful values of $\lambda$ and $\gamma_m$. From these plots, we can conclude that the critical point $E$ is a \textit{spiral sink} and \textit{asymptotically stable} in nature for $(\lambda=\pm2,\gamma_m=1)$, $\left(\lambda=\pm \sqrt{6},\gamma_m=\frac{4}{3}\right)$.

\begin{table}[h]
	\caption{\label{T2}Table shows the stability of each critical point for both hyperbolic and nonhyperbolic cases corresponding to every critical point:}
	\begin{tabular}{|c|c|c|}
		\hline
	\begin{tabular}{@{}c@{}} $~~$\\	 Critical Points  \\$~~$\end{tabular}  & \begin{tabular}{@{}c@{}} $~~$\\Stability for hyperbolic case \\$~~$\end{tabular} & \begin{tabular}{@{}c@{}} $~~$\\Stability for nonhyperbolic case  \\$~~$\end{tabular}\\ \hline\hline
		$A$  & \begin{tabular}{@{}c@{}} $~~$\\Saddle point for $\frac{2}{3}<\gamma_m<2$,\\ source for 
$0<\gamma_m<\frac{2}{3}$\\$~~$\end{tabular} &\begin{tabular}{@{}c@{}} $~~$\\Asymptotically stable for $\gamma_m=0$,\\ unstable for $\gamma_m=\frac{2}{3}$\\$~~$\end{tabular}\\ \hline
$B$  & \begin{tabular}{@{}c@{}}$~~$\\ Source for $\lambda<\sqrt{6}$ or $\lambda>\frac{5\sqrt{2}}{\sqrt{3}}$, \\ saddle point for $\lambda\in\left(\sqrt{6},\frac{5\sqrt{2}}{\sqrt{3}}\right)$
\\$~~$\end{tabular}& \begin{tabular}{@{}c@{}}$~~$\\ Semistable for $\lambda=\sqrt{6}$ and all $\gamma_m$,\\ semistable for $\lambda=\frac{5\sqrt{2}}{\sqrt{3}}$ and $\frac{2}{3}<\gamma_m<2$,\\ unstable for $\lambda=\frac{5\sqrt{2}}{\sqrt{3}}$ and $\gamma_m<\frac{2}{3}$\\$~~$\end{tabular} \\ \hline
$C$  & \begin{tabular}{@{}c@{}}$~~$\\ Source for $\lambda>-\sqrt{6}$ or $\lambda<-\frac{5\sqrt{2}}{\sqrt{3}}$, \\ saddle point for $\lambda\in \left(-\frac{5\sqrt{2}}{\sqrt{3}},-\sqrt{6}\right)$
\\$~~$\end{tabular} & \begin{tabular}{@{}c@{}}$~~$\\ Semistable for $\lambda=-\sqrt{6}$ and all $\gamma_m$,\\ asymptotically stable for $\lambda=-\frac{5\sqrt{2}}{\sqrt{3}}$\\ and $\frac{2}{3}<\gamma_m<2$,\\ unstable for $\lambda=-\frac{5\sqrt{2}}{\sqrt{3}}$ and $\gamma_m<\frac{2}{3}$\\$~~$\end{tabular} \\ \hline
$D$  & \begin{tabular}{@{}c@{}}$~~$\\ Stable node for $\lambda^2<3\gamma_m$,\\saddle point for $3\gamma_m<\lambda^2<6$, \\
unstable node for $\lambda^2>6$\\$~~$
\end{tabular} & \begin{tabular}{@{}c@{}}$~~$\\ 
Asymptotically stable for $\lambda=\sqrt{2}$\\ and $\gamma_m<0.342$ or $\gamma_m>\frac{2}{3}$, \\ unstable for $\lambda=\sqrt{2}$ and $0.342<\gamma_m<\frac{2}{3}$,\\ asymptotically stable for $\lambda=-\sqrt{2}$ \\ and $\gamma_m>1.245$ or $\gamma_m<\frac{2}{3}$, \\ unstable for $\lambda=-\sqrt{2}$ and $\frac{2}{3}<\gamma_m<1.245$,  \\ asymptotically stable for $\gamma_m=\frac{\lambda^2+2}{3}$ \\where $\lambda\in (-\sqrt{6},\sqrt{6})$
\\$~~$
\end{tabular}  \\ \hline\begin{tabular}{@{}c@{}}$~~$\\ $E$ \\$~$\end{tabular} & \begin{tabular}{@{}c@{}}$~~$\\Spiral sink for $\lambda=\pm 2,\gamma_m=1$ \\and $\lambda=\pm \sqrt{6},\gamma_m=\frac{4}{3}$.\\$~$\end{tabular}& \begin{tabular}{@{}c@{}}$~~$\\Very complicated to analyze.\\$~$\end{tabular} \\ \hline 
	\end{tabular}	
\end{table}

\section{Cosmological Implications}\label{IV}

\begin{figure}[h]
	\centering
	\begin{subfigure}[b]{0.475\textwidth}
		\centering
		\includegraphics[width=\textwidth]{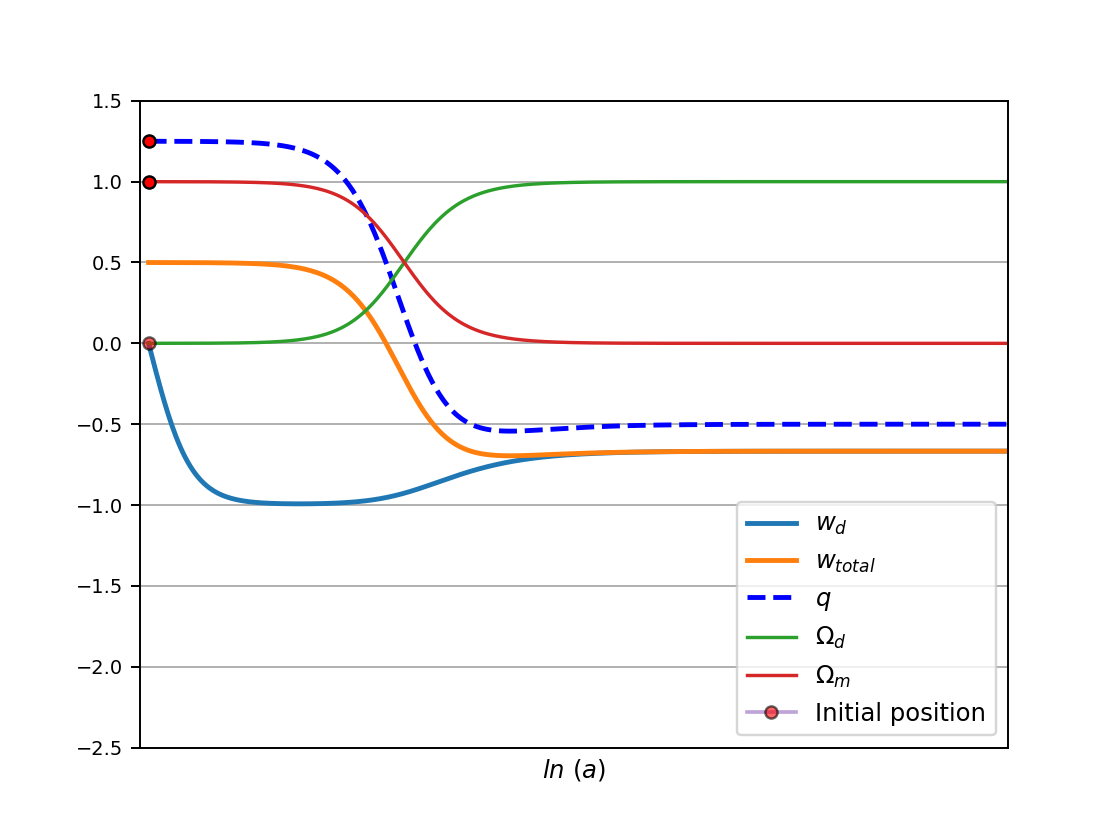}
		\caption{}
	\end{subfigure}
	\quad
	\begin{subfigure}[b]{0.475\textwidth}  
		\centering 
		\includegraphics[width=\textwidth]{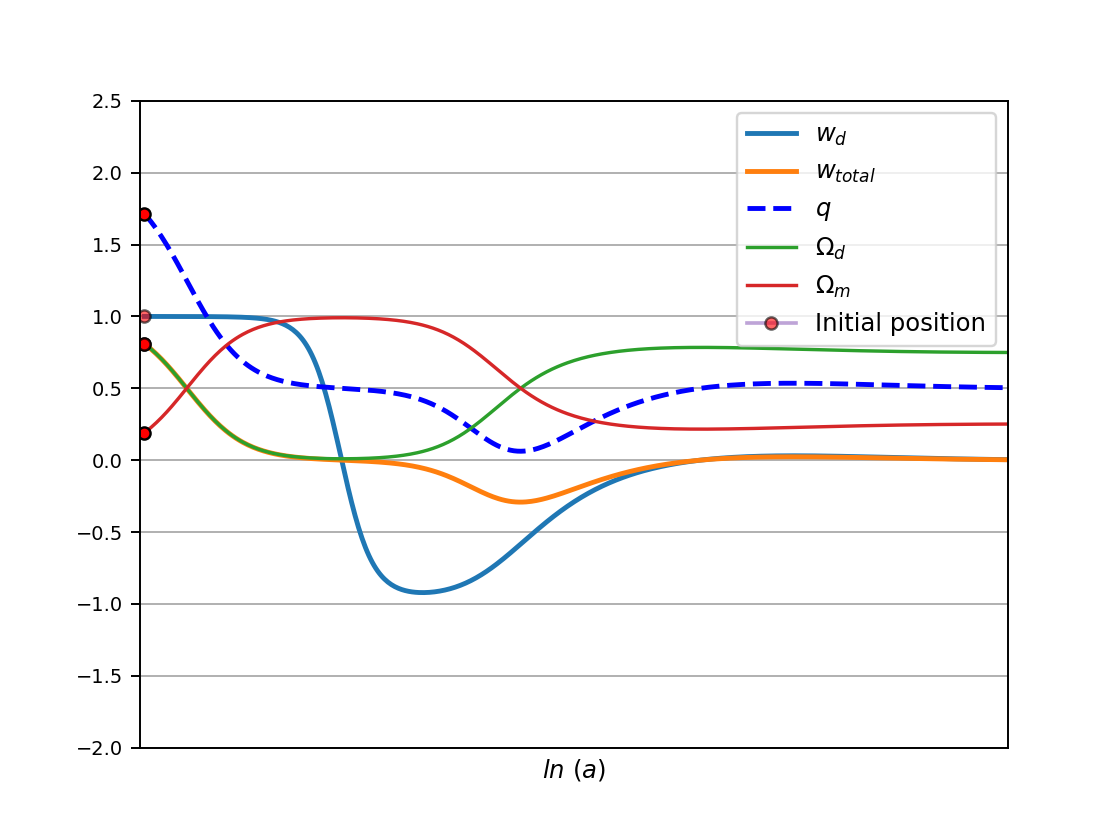}
		\caption{}
	\end{subfigure}
	\vskip\baselineskip
	\begin{subfigure}[b]{0.475\textwidth}   
		\centering 
		\includegraphics[width=\textwidth]{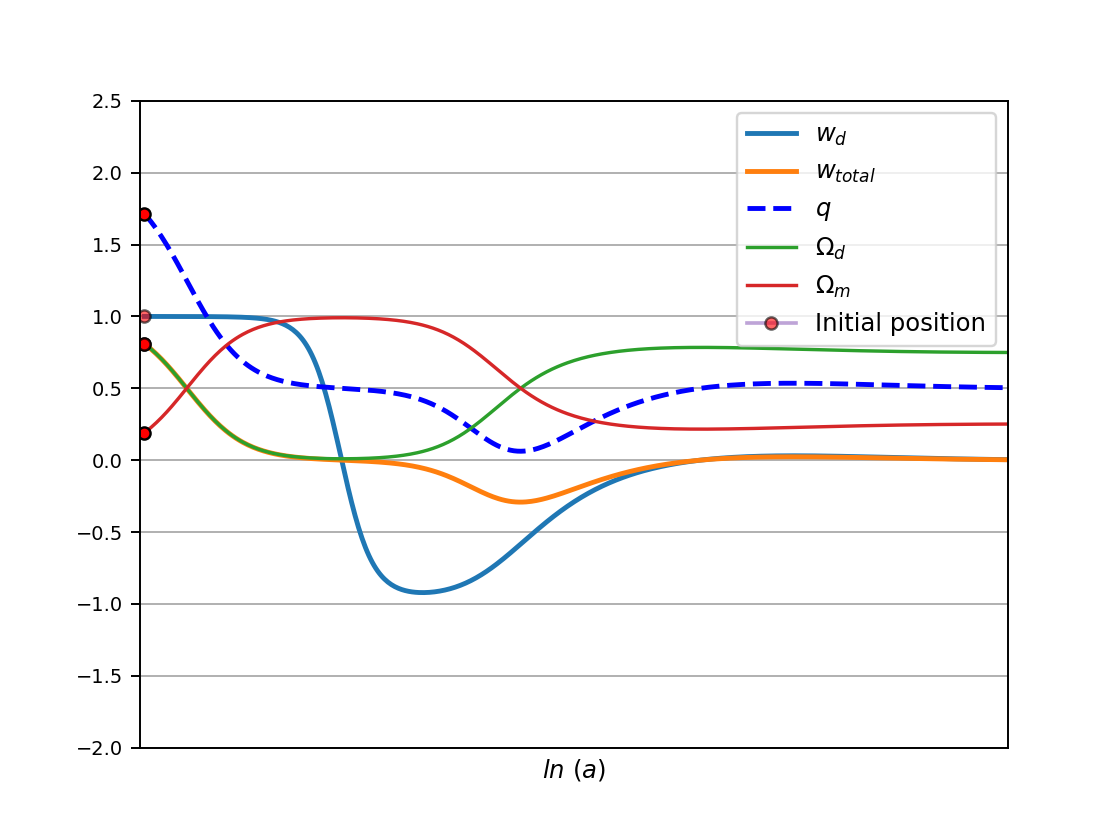}
		\caption{}
	\end{subfigure}
	\quad
	\begin{subfigure}[b]{0.475\textwidth}   
		\centering 
		\includegraphics[width=\textwidth]{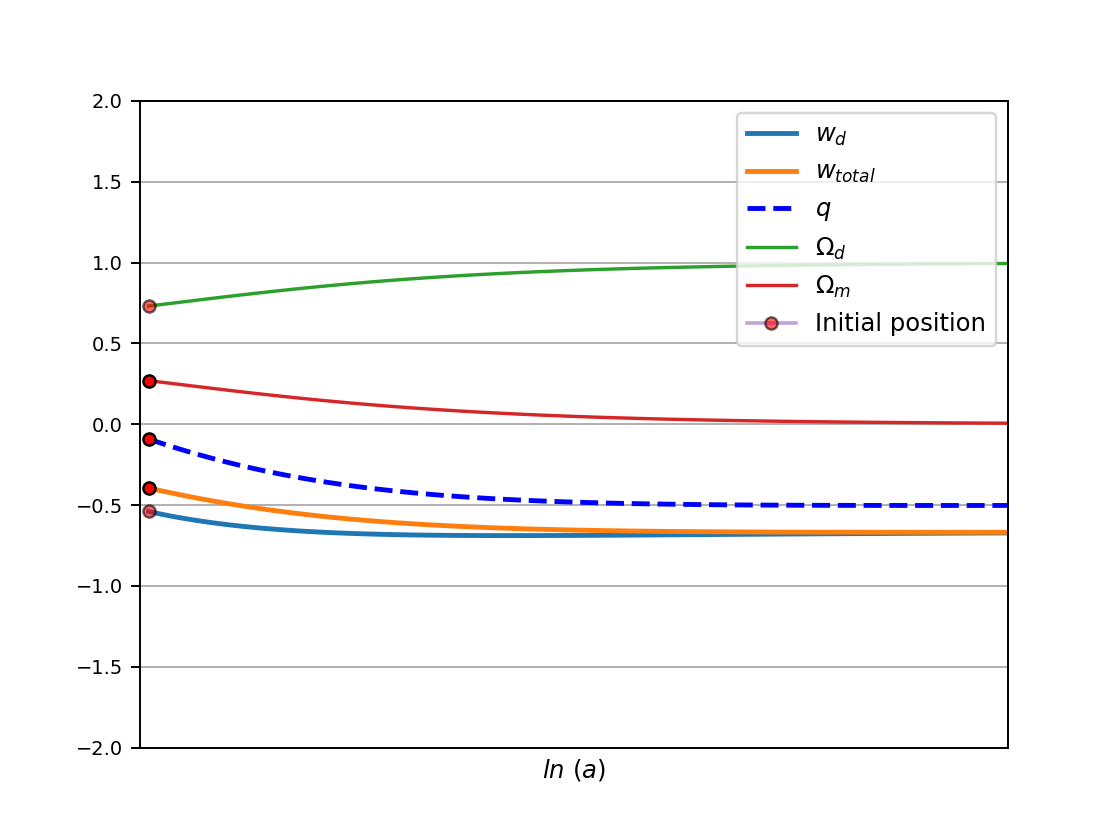}
		\caption{}
	\end{subfigure}
	\caption{The figure shows the time evolution of cosmological parameters of our cosmological model. In panel
		(a), initial position near the critical point $A$, for $\lambda=1$ and $\gamma_m=1.5$ the kinetic energy dominated late time accelerated solutions attracted towards quintessence era. In panel (b) and (c), initial position near the critical points $B$ and $C$ respectively, for $\lambda=2$ and $\gamma_m=1$ we get the kinetic energy and dust dominated late-time decelerated solutions. In panel (d), the behavior of the late solution near the critical point $D$ is similar to panel (a).}
\end{figure}
It is to be noted that the constraint equation $\Omega_{\phi}=x^2+ y^2$ yields $0\leq x^2+y^2 \leq 1$, for a non-negative fluid density $\rho_d \geq 0$.  So the evolution of this system is completely confined to a unit disc.  The lower half-disc i.e. $y<0$ transpires $H<0$ which corresponds to contracting universes.  Without loss of generality, we restrict our discussion to the upper-half disc ($y \geq 0$) due to the symmetry of the system under reflection and time reversal.\\

 The fixed point ($x=0,y=0$) corresponds to the dark matter dominated solution where $\Omega_d =0$ is saddle for $\gamma_m \in \left(\frac{2}{3},\frac{4}{3}\right)\cup \left(\frac{4}{3}, 2\right)$, source for $\gamma_m \in (0,\frac{2}{3})$ and unstable in nature at $\gamma_m=\frac{2}{3}$.  The matter-dominated solution is stable for $\omega_m=-1$.  Two  of the fixed points ($x=\pm 1, y=0$) correspond to scalar field-dominated solutions where $\omega_d=1$ is unstable in nature. For $\lambda=-\frac{5\sqrt{2}}{\sqrt{3}}$, the fixed point $C$ is asymptotically stable  for $\frac{2}{3} < \gamma_m < 2$.  On the other hand, the fixed point $D$ is corresponds to scalar field dominated solution ($\Omega_d =1$) which exists for $-\sqrt{6} \leq \lambda \leq \sqrt{6}$. $\omega_d =\frac{\lambda ^2}{3}-1$ gives rise to a power law inflationary expansion ($q<0)$ for $\lambda ^2 <2$.  We have shown that $D$ is a stable node while $\lambda ^2 <3 \gamma_m$ and hyperbolic fixed point.
 For non-hyperbolic case: for $\lambda = \sqrt{2}$, $D$ is asymptotically stable when $\gamma_m < 0.342$ or $\gamma_m>\frac{2}{3}$. For $\lambda= -\sqrt{2}$, $D$ is asymptotically stable when $\gamma_m>1.245$ or $\gamma_m < \frac{2}{3}$.  For $\lambda \in (-\sqrt{6},\sqrt{6})$ the point $D$ is asymptotically stable when $\gamma_m=\frac{\lambda^2 +2}{3}$.\\

Barotropic fluid dominated solution $(x=0, y=0)$ is hyperbolic where $\Omega_d=0$ is saddle in nature $\frac{2}{3}< \gamma_m < 2$ source for $0<\gamma_m < \frac{2}{3}$. For $\gamma_m =\frac{2}{3}$ and $0$, this solution is unstable and asymptotically stable accordingly and non-hyperbolic in nature. \\

Two of the fixed points $(x=\pm 1, y=0)$ correspond to solutions where the constraint equation (\ref{FME2a}) is dominated by the kinetic energy of the scalar field with a stiff equation of state, $\Omega_d=1$.  The fixed point $B$ is hyperbolic in nature and source for $\lambda<\sqrt{6}$ or $\lambda>\frac{5\sqrt{2}}{\sqrt{3}}$, and saddle for $\lambda\in\left(\sqrt{6},\frac{4\sqrt{2}}{\sqrt{3}}\right)\cup\left(\frac{4\sqrt{2}}{\sqrt{3}},\frac{5\sqrt{2}}{\sqrt{3}}\right)$.  $B$ is non-hyperbolic and semistable for $\lambda=\sqrt{6}, \frac{5\sqrt{2}}{\sqrt{3}}$ and $\frac{2}{3}< \gamma_m < 2$ and unstable for $\lambda = \frac{5\sqrt{2}}{\sqrt{3}}$ and $\gamma_m<\frac{2}{3}$. \\

The fixed point $C$ is hyperbolic in nature and source for $\lambda\in(-\infty,-\frac{5\sqrt{2}}{\sqrt{3}})\cup(-\sqrt{6},\infty)$, and  saddle point for $\lambda\in\left(-\frac{4\sqrt{2}}{\sqrt{3}},-\sqrt{6}\right)\cup\left(-\frac{5\sqrt{2}}{\sqrt{3}},-\frac{4\sqrt{2}}{\sqrt{3}}\right)$.  $C$ is non-hyperbolic and semistable for $\lambda=-\sqrt{6}$, asymptotically stable for $\lambda=\frac{-5\sqrt{2}}{\sqrt{3}}$.  $C$ is non-hyperbolic and semistable for $\lambda=\frac{-5\sqrt{2}}{\sqrt{3}}$ and $\frac{2}{3}< \gamma_m <2$, unstable for $\lambda = \frac{-5\sqrt{2}}{\sqrt{3}}$ and $\gamma_m=\frac{2}{3}$.\\

The critical point $D$ corresponds to a scalar field-dominated solution is hyperbolic in nature and late-time attractor for $\lambda ^2 <3 \gamma_m$, saddle for $3\gamma_m <\lambda^2 < 6$, and unstable node for $\lambda^2>6$.  On the other hand, $D$ is non-hyperbolic in nature and corresponds to late time attractor for $\lambda=\sqrt{2}$ and $0.342<\gamma_m<\frac{2}{3}$, late time attracting solution for $\lambda=-\sqrt{2}$ and $\gamma_m> 1.245$ or $\gamma_m =\frac{2}{3}$, unstable for $\lambda =-\sqrt{2}$ and $\frac{2}{3}<\gamma_m <1.245$, late time attracting solution for $\gamma_m=\frac{\lambda^2+2}{3}$, where $\lambda \in (-\sqrt{6}, \sqrt{6})$. 
$E$ corresponds to scaling solution.  $E$ is hyperbolic in nature and spiral sink for $\lambda=\pm 2$, $\gamma=1$ and $\lambda=\pm \sqrt{6}, \gamma_m=\frac{4}{3}$.
We can divide the evolutionary character of the Universe into two groups: (a) generic evolution (b) non-generic evolution.  For $\gamma_m=0$, generic evolution occurs near the point $A$ as the family of orbits emerges from a stable equilibrium in the past and then finishes towards $A$ which represents the $\Lambda$CDM era.  And for $0<\gamma_m < \frac{2}{3}$, a generic evolution occurs near $A$ which represents the quintessence state as the orbit emerges from an unstable equilibrium point $A$ in the past and finishes at $D$.

\section{Brief Discussion and Concluding Remarks}\label{V}

The present work is an example of a discrete dynamical system in cosmology.  In a particular cosmological study of the scalar field, cosmology has been analyzed using a dynamical system at the background level.  For simplicity of the formation of the dynamical system, the scalar field potential is chosen in the exponential form.  At the background level, there are five critical points for which the resulting equation of state parameter of the resulting total fluid has been given in the last but one column of Table \ref{T1}.  The value of $\omega_{total}$ for the fixed points $B$ and $C$ indicates that the resulting fluid is a stiff fluid which is very important at the very early era of evolution, particularly when the quantum effects are important.  The fixed points $A$ and $E$ will corresponds to dark energy model if $\gamma_m<\frac{2}{3}$, while the resulting fluid will be a normal fluid if $\gamma_m>\frac{2}{3}$.  Similarly, the fixed point $D$ corresponds to a dark energy fluid if $\lambda^2<2$.  Thus the critical points $A$, $D$, and $E$ may have dark energy fluid as the resulting fluid with proper choice of the parameter involved.  Finally, we have the following observations from the perspective of stability.  For the critical points $A$ and $E$, there will be matter dominated era when $\gamma_m\in \left(\frac{2}{3},2\right)$ and it is a stable configuration for $\gamma_m=0$ and unstable when $\gamma_m\neq 0$.  The critical point $B$ is unstable for all possible  choices of $\lambda$ whereas the fixed point $C$ will be stable if $\frac{2}{3}<\gamma_m<2, ~\lambda=-\frac{5\sqrt{2}}{\sqrt{3}}$ and it will be unstable for other possible values of $\lambda$ and $\gamma_m$.  Lastly, the critical point $D$ is stable for $\lambda^2<3\gamma_m$ and unstable for other possible cases where $\gamma_m\in [0,2)$.  Further, when $\lambda=\sqrt{2}$, the critical point is stable when $\gamma_m<0.342$ or $\gamma_m>\frac{2}{3}$ and similarly when $\lambda=-\sqrt{2}$, the fixed point will be stable for $\gamma_m>1.245$ or $\gamma_m<\frac{2}{3}$.  However, the model will be stable if $\gamma_m=\frac{\lambda^2+2}{3}$ and $\lambda\in (-\sqrt{6},\sqrt{6})$.	Finally, the satbility analysis shows the matter dominated phase or the present accelerated expansion of the universe depending on the  choices of the appropriate parameters and they may corresponds to stable or unstable configuration depending on the restrictions on the parameter.

	\smallbreak

\begin{acknowledgements}
The authors would like to thank the reviewer whose comments and suggestions improved the quality of the manuscript.	The author Soumya Chakraborty is grateful to CSIR, Govt. of India for giving Senior Research Fellowship (CSIR Award No: 09/096(1009)/2020-EMR-I) for the Ph.D. work.\\
\end{acknowledgements}

\noindent \textbf{Conflict of Interest}

There is no conflict of interest for this paper.  The authors declare the following financial interests/personal relationships which may be considered as potential competing interests: Soumya Chakraborty reports financial support was provided by Council of Scientific \& Industrial research, India.\\

\noindent \textbf{Data availability} 

This article describes entirely theoretical research. So data sharing is not applicable to this article as no datasets were generated or analyzed during the current study.\\\\

\bibliographystyle{unsrt}

\bibliography{ref}

\begin{thebibliography}{10}

\bibitem{BAHAMONDE20181}
Sebastian Bahamonde, Christian~G. Böhmer, Sante Carloni, Edmund~J. Copeland,
  Wei Fang, and Nicola Tamanini.
\newblock Dynamical systems applied to cosmology: Dark energy and modified
  gravity.
\newblock {\em Physics Reports}, 775-777:1--122, 2018.

\bibitem{wainwright_ellis_1997}
{\em Dynamical Systems in Cosmology}.
\newblock Cambridge University Press, 1997.

\bibitem{Sangwan:2018zpz}
Archana Sangwan, Ashutosh Tripathi, and H.~K. Jassal.
\newblock {Observational constraints on quintessence models of dark energy}.
\newblock 4 2018.

\bibitem{Liu:2022yao}
Jackie C.~H. Liu.
\newblock {A quintessence dynamical dark energy model from ratio gravity}.
\newblock {\em Eur. Phys. J. C}, 82(2):165, 2022.

\bibitem{Bouali:2023ncq}
Amine Bouali, Himanshu Chaudhary, Amritansh Mehrotra, and S.~K.~J. Pacif.
\newblock {Model-independent study for a quintessence model of dark energy:
  Analysis and Observational constraints}.
\newblock 4 2023.

\bibitem{article}
Savas Arapoglu and A.~Emrah Yükselci.
\newblock Dynamical system analysis of quintessence models with exponential
  potential - revisited.
\newblock {\em Modern Physics Letters A}, 34, 11 2017.

\bibitem{Goswami:2019zto}
Gopi~Kant Goswami, Anirudh Pradhan, and Aroonkumar Beesham.
\newblock {A Dark Energy Quintessence Model of the Universe}.
\newblock {\em Mod. Phys. Lett. A}, 35(04):2050002, 2019.

\bibitem{Oikonomou:2017ppp}
V.~K. Oikonomou.
\newblock {Autonomous dynamical system approach for inflationary
  Gauss\textendash{}Bonnet modified gravity}.
\newblock {\em Int. J. Mod. Phys. D}, 27(05):1850059, 2018.

\bibitem{Odintsov:2017tbc}
S.~D. Odintsov and V.~K. Oikonomou.
\newblock {Autonomous dynamical system approach for $f(R)$ gravity}.
\newblock {\em Phys. Rev. D}, 96(10):104049, 2017.

\bibitem{Chakraborty:2021pkp}
Soumya Chakraborty, Sudip Mishra, and Subenoy Chakraborty.
\newblock {Dynamical system analysis of self-interacting three-form field
  cosmological model: stability and bifurcation}.
\newblock {\em Eur. Phys. J. C}, 81(5):439, 2021.

\bibitem{Chakraborty:2020vkp}
Soumya Chakraborty, Sudip Mishra, and Subenoy Chakraborty.
\newblock {Dynamical system analysis of three-form field dark energy model with
  baryonic matter}.
\newblock {\em Eur. Phys. J. C}, 80(9):852, 2020.

\bibitem{JAWAD2022101127}
Abdul Jawad and Abdul~Malik Sultan.
\newblock Analyzing stability of five-dimensional einstein chern–simons
  gravity through dynamical systems.
\newblock {\em Physics of the Dark Universe}, 38:101127, 2022.

\bibitem{CAPOZZIELLO2022137475}
Salvatore Capozziello, Rocco D'Agostino, and Orlando Luongo.
\newblock The phase-space view of non-local gravity cosmology.
\newblock {\em Physics Letters B}, 834:137475, 2022.

\bibitem{ACQUAVIVA2022101128}
Giovanni Acquaviva and Nihan Katırcı.
\newblock Dynamical analysis of logarithmic energy–momentum squared gravity.
\newblock {\em Physics of the Dark Universe}, 38:101128, 2022.

\bibitem{2022arXiv220206648B}
Z.~S. {Boxonov}.
\newblock {A discrete-time dynamical system of mosquito population}.
\newblock {\em arXiv e-prints}, page arXiv:2202.06648, February 2022.

\bibitem{Alho_2022}
Artur Alho, Woei~Chet Lim, and Claes Uggla.
\newblock Cosmological global dynamical systems analysis.
\newblock {\em Classical and Quantum Gravity}, 39(14):145010, jun 2022.

\bibitem{Humieja:2019ywy}
Franciszek Humieja and Marek Szyd\l{}owski.
\newblock {Bifurcations in Ratra\textendash{}Peebles quintessence models and
  their extensions}.
\newblock {\em Eur. Phys. J. C}, 79(9):794, 2019.

\bibitem{Feng:2014fsa}
Chao-Jun Feng, Xin-Zhou Li, and Li-Yan Liu.
\newblock {Bifurcation and global dynamical behavior of the $f$(T) theory}.
\newblock {\em Mod. Phys. Lett. A}, 29(07):1450033, 2014.

\bibitem{Mishra:2021sza}
Sudip Mishra.
\newblock {\em {Dynamics around nonhyperbolic equilibrium and bifurcation
  analysis of various cosmological models}}.
\newblock PhD thesis, Jadavpur University, Department of Mathematics, India,
  2021.

\bibitem{Biswas:2016ces}
Sujay~Kr. Biswas and Subenoy Chakraborty.
\newblock {Interacting dark energy model in the brane scenario: A Dynamical
  System Analysis}.
\newblock {\em Int. J. Geom. Meth. Mod. Phys.}, 16(08):1950115, 2019.

\bibitem{article3}
Nandan Roy and Narayan Banerjee.
\newblock Quintessence scalar field: A dynamical systems study.
\newblock {\em The European Physical Journal Plus}, 129, 02 2014.

\bibitem{Singh:2019enu}
S.~Surendra Singh and Chingtham Sonia.
\newblock {Dynamical system perspective of cosmological models minimally
  coupled with scalar field}.
\newblock {\em Adv. High Energy Phys.}, 2020:1805350, 2020.

\bibitem{book:1378164}
Saber~N Elaydi.
\newblock {\em Discrete Chaos, Second Edition : With Applications in Science
  and Engineering}.
\newblock CRC Press, 2nd ed edition, 2007.

\bibitem{BISWAS2020145}
Milan Biswas and Nandadulal Bairagi.
\newblock On the dynamic consistency of a two-species competitive discrete
  system with toxicity: Local and global analysis.
\newblock {\em Journal of Computational and Applied Mathematics}, 363:145--155,
  2020.

\bibitem{Saha2018}
Priyanka Saha, Nandadulal Bairagi, and Milan Biswas.
\newblock {\em On the Dynamics of a Discrete Predator--Prey Model}, pages
  219--232.
\newblock Springer International Publishing, Cham, 2018.

\end{thebibliography}

\bigbreak 

\appendix

\section{Criteria for stability}\label{A}

\subsection{Hyperbolic Fixed Points}	
	
The following result is the main tool for detecting local stability.
\begin{theorem}
	Let $x^*$ be a hyperbolic fixed point of a map $f$, where $f$ is continuously differentiable at $x^*$. The following statements then hold true \cite{book:1378164}:
	$$
	\begin{aligned}
	 1. & ~ \text{If $|f'(x^*)|<1$, then $x^*$ is asymptotically stable.}\\
	 2. & ~ \text{If $|f'(x^*)|>1$, then $x^*$ is unstable.}
	\end{aligned}
	$$
	\end{theorem}

	\subsection{Nonhyperbolic Fixed Points}
	
	The stability criteria for nonhyperbolic fixed points are more involved. They
	will be summarized in the next two results, the first of which treats the case
	when $f'(x^*)=1$ and the second for $f'(x^*)=-1$.
	
	\begin{theorem}\label{2}
		Let $x^*$ be a fixed point of a map $f$ such that $f'(x^*)=1$. If $f'(x)$, $f''(x)$, and $f'''(x)$ are continuous at $x^*$, then the following statements hold \cite{book:1378164}:
		$$
		\begin{aligned}
		1. & ~ \text{If $f''(x^*)\neq 0$, then $x^*$ is unstable (semistable).}\\
		2. & ~ \text{If $f''(x^*)=0$ and $f'''(x^*)<0$, then $x^*$ is aymptotically stable.}\\
		3. & ~ \text{If $f''(x^*)=0$ and $f'''(x^*)>0$, then $x^*$ is unstable.}
		\end{aligned}
		$$
	\end{theorem}
	
	The preceding theorem may be used to establish stability criteria for the case when $f'(x^*)=-1$. But before doing so, we need to introduce the notion of the Schwarzian derivative.

	\begin{definition}
		The Schwarzian derivative, $Sf$, of a function $f$ is defined by \cite{book:1378164}
		$$
		Sf(x)=\frac{f'''(x)}{f'(x)}-\frac{3}{2} \left[\frac{f''(x)}{f'(x)}\right]^2 .
		$$
		And if $f'(x^*)=-1$, then
		$$
		Sf(x^*)=-f'''(x^*)-\frac{3}{2}\left[f''(x^*)\right]^2.
		$$
	\end{definition}

		\begin{theorem}\label{3}
		Let $x^*$ be a fixed point of a map $f$ such that $f'(x^*)=-1$. If $f'(x)$, $f''(x)$, and $f'''(x)$ are continuous at $x^*$, then the following statements hold \cite{book:1378164}:
		$$
		\begin{aligned}
		1. & ~ \text{If $Sf(x^*)<0$, then $x^*$ is asymptotically stable.}\\
		2. & ~ \text{If $Sf(x^*)>0$, then $x^*$ is unstable.}\\
		\end{aligned}
		$$
	\end{theorem}

		\begin{theorem}
		Let $f:G\subset \mathbb{R}^2 \rightarrow \mathbb{R}^2$ be a $C^1$ map, where $G$ is an open subset of $\mathbb{R}^2$, $X^*$ is a fixed point of $f$, and $A=Df(X^*)$.  Let $\rho(A)$ be the spectral radius of $A$ defined as $\rho(A)=\max \{|\lambda_1|, |\lambda_2|: \lambda_1, \lambda_2 ~\text{are the eigen values of $A$}\}$.  Then the following statements hold true:
		$$
		\begin{aligned}
			1. & ~ \text{If $\rho(A)<1$, then $X^*$ is asymptotically stable.} \\
			2. & ~ \text{If $\rho(A)>1$, then $X^*$ is unstable.}\\
			3.&~\text{If $\rho(A)=1$, then $X^*$ may or may not be stable.}
		\end{aligned}
		$$
	\end{theorem}
	
\subsection{Center Manifolds}

Consider the $s$-parameter map $F(\mu,u)$, $F:\mathbb{R}^s\times \mathbb{R}^k\rightarrow \mathbb{R}^k$, with $u\in \mathbb{R}^k$, $\mu \in \mathbb{R}^s$, where $F$ is $C^r$ $(r\geq 3)$ on some sufficiently large open set in $\mathbb{R}^k\times \mathbb{R}^s$.  Let $(u^*,\mu^*)$ be a fixed point of $F$, i.e., 
$$
F(\mu^*,u^*)=u^*.
$$
The stability of hyperbolic fixed points of $F$ is determined from the stability of the fixed points under the linear map
\begin{align}
J=D_\mu F(\mu^*,u^*). \label{Jacobian}
\end{align}
However, the situation is drastically different if one of the eigenvalues $\lambda$ of $J$ lies on the unit circle, that is, $|\lambda|=1$.  There are three separate cases in which	the fixed point $(u_0,\mu_0)$ is non-hyperbolic.
$$
\begin{aligned}
1.&~ \text{$J$ has one real eigenvalue equal to 1 and the other eigenvalues are off	the unit circle.} \\
2.&~ \text{$J$ has one real eigenvalue equal to $-1$ and the other eigenvalues are off the unit circle.}\\
3.&~\text{$J$ has two complex conjugate eigenvalues with modulus $1$ and the other eigenvalues are off the unit circle.}
\end{aligned}
$$
Let us temporarily suppress the parameter $\mu$.  Then the map $F$ can be written in the form
\begin{equation}
\begin{aligned}
x & \longmapsto Ax+f(x,y), \\  y & \longmapsto By+g(x,y)
\end{aligned}\label{system}
\end{equation}
where $J$ in equation $(\ref{Jacobian})$ has the form $J=\begin{pmatrix} A&0\\0&B\end{pmatrix}$.  Moreover, all of the eigenvalues of $A$ lie on the unit circle, and all of the eigenvalues of $B$ are off the unit circle. Furthermore,
$$
\begin{aligned}
& f(0,0)=0,\quad g(0,0)=0, \\ & Df(0,0)=0, \quad Dg(0,0)=0.
\end{aligned}
$$

\begin{theorem}
	There is a $C^r$-center manifold for system $(\ref{system})$ that can be represented locally as \cite{book:1378164}
	$$
	M_c=\left\{(x,y)\in \mathbb{R}^t\times \mathbb{R}^s : y=h(x),|x|<\delta,h(0)=0,Dh(0)=0,~\text{for a sufficiently small}~ \delta  \right\}.
	$$
	Furthermore, the dynamics restricted to $M_c$ are given locally by the map
	\begin{align}
	x \longmapsto Ax+f(x,h(x)), x\in \mathbb{R}^t.\label{cm}
	\end{align}
\end{theorem}	
This theorem asserts the existence of a center manifold, i.e., a curve $y=h(x)$ on which the dynamics of System (\ref{system}) is given by Equation (\ref{cm}). The next result states that the dynamics on the center manifold $M_c$ determines completely the dynamics of the System (\ref{system}).
\begin{theorem}\label{7}
	The following statements hold \cite{book:1378164}.\\
	
	1. If the fixed point $(0,0)$ of Equation (\ref{cm}) is stable, asymptotically stable,
	or unstable, then the fixed point $(0,0)$ of System (\ref{system}) is stable, asymptotically stable, or unstable, respectively.\\
	
	2. For any solution $(x(n),y(n))$ of the system (\ref{system}) with an initial point $(x_0,y_0)$ in a small neighborhood around the origin, there exists a solution $z(n)$ of Equation (\ref{cm}) and positive constants $L$, $\beta>1$ such that
	$$
	|x(n)-z(n)|\leq L\beta^n,~ \text{and}~ |y(n)-h(z(n))|\leq L\beta^n~\text{for all $n\in \mathbb{Z}^+$}.
	$$
	
\end{theorem}	 
	
The question that still lingers is how to find the center manifold $M_c$ or, equivalently, how to compute the curve $y=h(x)$. The first thing that comes
to mind is to substitute for $y$ in System (\ref{system}) to obtain the system
$$
\begin{aligned}
	x(n+1)&=Ax(n)+f(x(n),h(x(n))),\\
	y(n+1)&=h(x(n+1)) \\&=h\left[Ax(n)+f(x(n),h(x(n)))\right]\\&=Bh(x(n))+g(x(n),h(x(n))).
\end{aligned}
$$

Equating the two equations for $y(n + 1)$ yields the functional equation
\begin{align}
	\mathcal{F}(h(x))=h\left[Ax+f(x,h(x))\right]-Bh(x)-g(x,h(x))=0. \label{cal}
\end{align}
Solving Equation (\ref{cal}) is a formidable task, so at best one can hope to approximate its solution via power series. The next result provides the theoretical justification for our approximation.

\begin{theorem}\label{7A}
	Let $\psi : \mathbb{R}^t\rightarrow \mathbb{R}^s$ be a $C^{1}$-map with $\psi(0)=\psi'(0)=0$.  Suppose that $\mathcal{F}(\psi(x))=\mathcal{O}(|x|^r)$ as $x\rightarrow 0$  for some $r>1$ \cite{book:1378164}.  Then,
	$$
	h(x)=\psi(x)+\mathcal{O}(|x|^r)~\text{as}~x\rightarrow 0.
	$$
\end{theorem}

\section{Center manifold calculation for the critical point $A$ for case $\gamma_m=\frac{2}{3}$}\label{B}

For $\gamma_m=\frac{2}{3}$, the system $(\ref{eq3}-\ref{eq4})$ modifies to
$$
\begin{aligned}
\overline{x}&=-x+2 x^3-x y^2+\sqrt{\frac{3}{2}} \lambda  y^2, \\
\overline{y}&=2 y+2 x^2 y-\sqrt{\frac{3}{2}} \lambda  x y-y^3 .
\end{aligned}
$$
Consider the map $F=\begin{pmatrix}
	f\\g
\end{pmatrix}$ defined by
$$
\begin{pmatrix}
	x\\\\y
\end{pmatrix}\longmapsto  \begin{pmatrix}
	-1 & 0 \\ \\ 0 & 2
\end{pmatrix}\begin{pmatrix}
	x \\ \\y
\end{pmatrix}+\begin{pmatrix}
	2 x^3-x y^2+\sqrt{\frac{3}{2}} \lambda  y^2 \\
	2 x^2 y-\sqrt{\frac{3}{2}} \lambda  x y-y^3
\end{pmatrix}.
$$
Then,
$$
M_c=\left\{(x,y)\in \mathbb{R}^2 : y=h(x),h(0)=h'(0)=0\right\}.
$$

The function $h$ must satisfy Equation (\ref{cal})
$$
h\left(Ax+f(x,h(x))\right)-Bh(x)-g(x,h(x))=0
$$
or
\begin{align}
	h\left(-x+2x^3-xh^2(x)+\sqrt{\frac{3}{2}}\lambda h^2(x)\right)-2h(x)-\left(2x^2h(x)-\sqrt{\frac{3}{2}}\lambda x h(x)-h^3(x)\right)=0.\label{eq13}
\end{align}
Let us assume that $h(x)$ takes the form
\begin{align}
	h(x)=c_1 x^2+c_2 x^3+\mathcal{O}(x^4).\label{eq14}
\end{align}
Then, substituting Equation (\ref{eq14}) in Equation (\ref{eq13}) and writing the terms upto degree $3$ yields
$$
c_1x^2+c_2 x^3-2(c_1x^2+c_2x^3)+\sqrt{\frac{3}{2}}\lambda c_1 x^3+\cdots =0.
$$
As we analyze the arbitrary neighborhood of the origin so by comparing both sides the coefficient of $x^2$ and $x^3$ yields
$$
\begin{aligned}
	&	c_1-2c_1=0\implies c_1=0;\\
	&	c_2-2c_2+\sqrt{\frac{3}{2}}\lambda c_1=0 \implies c_2=0.
\end{aligned}
$$ 
Consequently, $h(x)=0$ and the map $f$ on the center manifold is given by
$$
x \longmapsto -x.
$$
It follows that the center manifold lies completely on the $x$ axis but we are unable to determine the flow on the center manifold.\\

	\end{document}